\documentclass[letterpaper,aps,prb,showpacs,twocolumn,groupedaddress]{revtex4-1}
\usepackage{graphics,color,epsfig}
\usepackage{amsmath,amssymb}

\def\e{\begin{equation}}
\def\f{\end{equation}}
\def\_#1{{\bf #1}}

\def\.{\cdot}

\def\E{\epsilon}

\newcommand{\ds}{\displaystyle}
\def\l#1{\label{eq:#1}}
\def\r#1{(\ref{eq:#1})}
\def\=#1{\overline{\overline #1}}
\def\##1{{\bf#1\mit}}

\renewcommand{\Re}{\mathop{\rm Re}}
\newcommand{\ka}{\hbox{\ae}}

\begin{document}

\title{Phase matched backward-wave second harmonic generation in a hyperbolic carbon nanoforest}

\author{A. K. Popov$^{1}$, I. S. Nefedov$^{2,3}$, S. A. Myslivets$^{4,5}$}
\affiliation{$^1$Birck Nanotechnology Center, Purdue University,
West Lafayette, IN 47907,~USA\\
$^2$Aalto University, School of Electrical Engineering,
P.O. Box 13000, 00076 Aalto, Finland\\
$^3$Laboratory Nanooptomechanics, ITMO University, St. Petersburg, 197101, Russia\\
$^4$L. V. Kirensky Institute of Physics, Siberian Division of the Russian Academy of Sciences, 660036 Krasnoyarsk, Russia \\
$^5$Siberian Federal University, 79 Svobodny pr., 660041 Krasnoyarsk, Russia}


\date{}


\begin{abstract}%

We show that  deliberately engineered spatially dispersive metamaterial slab can enable co-existence and phase matching of contra-propagating  ordinary fundamental and backward second harmonic electromagnetic modes. Energy flux and phase velocity are contra-directed in backward waves which determines extraordinary nonlinear-optical propagation processes. Frequencies, phase and group velocities, as well as nanowavequide losses inherent to the  electromagnetic modes supported by the metamaterial can be tailored to optimize nonlinear-optical conversion of frequencies and propagation directions of the coupled waves. Such a possibility, which is of paramount importance for nonlinear photonics, is proved with numerical model of the hyperbolic metamaterial made of carbon nanotubes standing on metal surface. Extraordinary properties of backward-wave second harmonic in the THz and IR propagating in the reflection direction  are investigated with  focus on  pulsed regime.
\end{abstract}

\pacs{41.20.Jb, 42.25.Bs, 42.65.Ky, 42.65.Sf, 77.84.Lf, 78.67.Ch, 78.67.Pt}
\maketitle

\section{Introduction}
{Metamaterials} (MM) are artificially designed and engineered materials, which can have properties unattainable in nature.
 MM rely on advances in nanotechnology to build tiny metallic nanostructures  smaller than the wavelength of light. These nanostructures modify the electromagnetic (EM) properties of MM, sometimes creating seemingly impossible optical effects.  Negative-index MM (NIM) are most intriguing EM materials that support  backward EM waves (BEMW). Phase velocity and energy flux (group velocity) become \emph{contra-directed} in NIM.  The appearance of BEMW  is commonly associated with simultaneously negative electric permittivity ($\epsilon<0$) and magnetic permeability ($\mu<0$) at the corresponding frequencies and, consequently,  to negative refractive index $n= - \sqrt{\mu\epsilon}$ [${k}^{2}=n^{2}(\omega/{c})^{2}$]. Such counter-intuitive backwardness of EMW  is in strict contrast with the electrodynamics of ordinary,  positive-index materials (PIM).
 It was generally accepted that the magnetization at optical frequencies is negligible \cite{Land} and, hence, did not play any essential role. It is because natural optical emitters (atoms) are much smaller than optical wavelengths. Consequently,
 magnetic permeability $\mu$ was normally set equal to one in the basic Maxwell's equations describing linear and NLO processes.
In the late 1960s, V.~G.~Veselago considered propagation of EM in an fictitious, isotropic medium with simultaneously negative dielectric permittivity $\epsilon$ and magnetic permeability $\mu$ and showed that it would exhibit unusual EM properties \cite{Vesel1,Vesel2}.
The situation has dramatically changed with the advent of nanotechnology. As was shown in \cite{Smith1,Smith11,Smith2,mu1,mu2,mu3,NIMExp1,NIMExp2}, specially shaped metallic nanostructures can enable negative optical magnetism.
While the physics and applications of  NIM are being explored world-wide at a rapid pace since then, current mainstream  still focuses on fabricating  of nanoscopic LC circuits  shaped as split rings, horse shoe, fishnets, etc. that ensures negative magnetic response. Size of such  mesoscopic structures (metaatoms and metamolecules) embedded in dielectric host material is typically about several tens to hundred nanometers. Since  excitation of electrons (plasma) localized in nanometer-scale  metallic building blocs  determines the outlined extraordinary EM properties of such  MM, they are commonly referred to as \emph{plasmonic} metamaterials (PMM).

Backwardness of EM waves dictates a revision of many concepts concerning EM propagation processes in man-made materials. This holds the promise of revolutionary breakthroughs in microwave and photonic device technologies \cite{Sh,MM}. Such breakthrough  can be employed to develop a wide variety of devices with enhanced and uncommon functions, such as enhanced photovoltaic light collection, biosensing, perfect imaging, all-optical memories and invisibility cloaks.   Nanostructured NIM are  expected to play a key role in the development of all-optical data processing chips.
Much success has already been demonstrated at microwave frequencies. This has led to commercially available products.
However, the situation is  different in the optical regime, where the visionary applications are expected to have the highest and most wide-ranging impact.

Nonlinear optical photonics in NIMs holds promises of the breakthrough that  essentially broaden the scope and power of the photonic applications \cite{ShC}.
It has been shown that opposite directions of the energy flow and phase velocity in NIM gives rise to unique  NLO propagation processes \cite{KivSHG,Sl,APB}.
The possibility to compensating losses inherent to most plasmonic MM  through coherent NLO energy exchange between ordinary and BEMW is among the important applications \cite{OL}. Such losses present a severe formidable obstacle towards  applications of NIMs. A review and corresponding references, including experimental realization, can be found in \cite{EPJD,SPIE,Mw,Such,OQE}. Creation of the MM that would enable {co-existence} of ordinary and BEMWs at the particular frequencies, their {phase matching} and strong NLO response is of paramount importance, while presents a great challenge.

 This paper is to study a novel paradigm for BW photonics that does not rely on optical magnetism \cite{APA2012,AST2013,SSP2014}.  At that, it enables  \emph{co-existence and phase matching} of ordinary and backward EMWs  through the \emph{alternative} approach  based on tailored variable quasi-hyperbolic dispersion. Such a possibility  and its general implementation principles are demonstrated with numerical simulations for  the particular example applicable to SHG in the THz and mid IR frequency ranges. It is shown that SHG in the proposed NIM exhibits exotic properties, especially in the pulse regime, that can be employed for unparallel applications to photonics and device technologies. Consideration of  particular technique for engineering of $\chi^{(2)}$ nonlinearity is beyond the scope of this work.

The basic idea is as follows. The limitations associated with characterization of NIM by negative $\epsilon$ and $\mu$ and the possibility of more general approach based on spatial inhomogeneity of \emph{dielectric} properties of materials at the nanoscale were  described in Refs. \cite{Agr,AgGa}.  The latter give rise to dispersion of the electromagnetic response of the medium to perturbations of frequency $\omega$ and wave vector $\mathbf{k}$. The presence of the spatial dispersion signifies a nonlocal dielectric response. It is manifested by the dependence of the generalized dielectric tensor $\epsilon_{ij}(\omega, \mathbf{k})$ on the wavevector $\mathbf{k}$ and may lead to  difference in signs of phase and group velocities.
  In a loss-free medium, energy flux $\mathbf{{S}}$  and directed along  group velocity ${\mathbf{v}_g}$ which may become directed \emph{against} the wavevector depending on the direction of the group velocity:
\begin{equation}
{\mathbf{S}}={\mathbf{v}_g}U,\quad 
{\bf v}_g={\rm grad}_{\bf k}\omega(\mathbf{k}), \label{gr}
\end{equation}
where $U$ is energy density. Hence, the metamaterial supports  BEMW modes  if dielectric properties and spatial dispersion of its structural elements lead to such relationship between the allowed frequencies and corresponding wavenumbers of polaritons  that $\partial\omega/\partial k<0$. The latter is  referred to as \emph{negative dispersion} of the medium electric response. The possibility to craft MM that can support BEMWs was shown in 2001 \cite{Tr}. In 2003 the authors of \cite{Smith} noticed the potential of
MM with a hyperbolic-type dispersion for subwavelength
imaging of objects at electrically large distances. The
concept of such imaging was first introduced in \cite{Pendry}.
The hyperbolic-type dispersion is inherent to uniaxial materials in which
the axial and tangential components of the permittivity and (or)
permeability tensors have different signs. Such media are referred to as { indefinite MM} \cite{Smith}.
The hyperbolic-type (or a conic-type in a 3D case) dispersion results
in a possibility to design a ``hyperlens'', where evanescent near
fields are transformed into propagating modes and can be transported
at electrically long distances \cite{Zhang,Shvets}. Thus, for a transfer of image with subwavelength resolution
media, characterized by the permittivity
tensors with negative signs of all components are not necessary.
 For waves of only one
linear polarization (TM-polarized waves with respect to the optical axis), that allows to realize a layer of an indefinite \emph{dielectric} MM \cite{Smith3}, whose
permeability is unity and only components of the permittivity tensor
have different signs. For the visible range, such a MM was
designed in \cite{Smith3,Kawata} as an array of parallel plasmonic (metal) nanowires.
Appearance of BEMW modes in the particular case of  layered structures has been shown  in Ref.~\cite{Bel,Nar}.  It was shown that nonlinear dielectrics add tunability to the optical response of layered hyperbolic metamaterials  \cite{Alu,Lapine}. The possibility to achieve phase matching in such structures was recently reported in Ref. \cite{Duncan}.

Obviously, many other options should have existed \cite{Mok,Chr}. Each of them leaves open questions, both fundamental and specific for each potential material and for its particular applications. Among them are  losses imposed by metal component which are different for different nanowaveguide modes.   In such a context, we elaborate here a particular realization of the  paradigm proposed in Ref. \cite{APA2012,AST2013,SSP2014} and the underlaying physical principles. Based on the outlined principles,  we show the possibilities of engineering of nonmagnetic metamaterial, which supports a set of modes with negative and sign-changing dispersion and enables the coexistence of both of BEMW and  ordinary modes that satisfy to phase matching required for second harmonic generation.  Losses at the corresponding frequencies are calculated and extraordinary  properties of BWSHG in such metaslab are investigated in the most important case of pulsed regime.  The results prove new avenue in engineering of nonlinear photonic metamaterials and ultra-miniature BW NLO photonic devices, such as frequency upconverting all-optically  controlled  metareflector  with unparalleled functional properties.

  \section{Backward electromagnetic waves in a metamaterial slab made of carbon nanotubes}

 As shown in Ref.  \cite{IgorPRB}, EM waves  in arrays of metallic carbon nanotubes (CNTs) posses a  hyperbolic dispersion and reasonably low losses in the THz and mid-IR ranges. The study was based on the impedance cylinder model and effective boundary conditions, applied  to individual CNTs, and Green's function which takes into account EM interaction between CNTs.
Simple effective medium model also was proposed for arrays of CNTs \cite{Minsk}.

Below, we show the hyperbolic-type dispersion enables co-existence of  both  forward \emph{and backward} waves in finite-thickness slabs of vertically standing CNTs and discuss benefits offered by these structures for nonlinear applications. It is proved that high-loss and narrow-band resonant structures, like split-ring resonators, exhibiting negative $\epsilon$ and $\mu$ are \emph{not required} for realization of \emph{backward wave} regime.
Proposed two-dimensional  periodic arrays of carbon nanotubes are fabricated
by many research groups.
Such CNT arrays form \emph{finite-thickness} slabs,
 which are used already as field emitters \cite{ShFan}, biosensors
\cite{YLin} and antennas \cite{Wang,DresselNat}.
We assume that all
nanotubes possess metallic properties.
\subsection{Impedance cylinder model of an individual carbon nanotube}
As an example,  single-wall zigzag CNTs will be considered that  possess metallic properties.
As a model of the individual metallic zigzag nanotube  an impedance cylinder will be used, which characterizes by complex dynamic conductivity and effective boundary conditions \cite{Slepyan}.
According to this model, a simple approximate expression for the complex surface conductivity can be employed which is valid for metallic zigzag CNTs at frequencies below optical transitions:
\e\l{a1} \sigma_{zz}\cong
i[{2\sqrt{3}e^2\Gamma_0}]/[{3q\pi\hbar^2(\omega+i2\pi\nu)}]. \f
Here
$e$ is electron charge, $\Gamma_0=2.7$~eV is the overlapping
integral, $\tau=1/\nu$ is the relaxation time, $\hbar$ is
reduced Planck constant. The radius of metallic zigzag CNT $r$ is determined
by an integer $q$ as $r=3\sqrt{3}qb/2\pi$, where $b=0.142$\,nm is
the interatomic distance in graphene.
Since the wall of a
single-wall CNT is a monoatomic sheet of carbon, Eq.~\r{a1} can be
considered as the surface conductivity of the carbon nanotube.
Simple formulas for the surface conductivity, like \r{a1}, were used by authors of many publications on EM properties of CNTs \cite{Hanson,Hanson2,BurkeNT,Maffucci,Miano,Mikki}.
The surface impedance per unit length is
\e\l{z}
z_i=\frac{1}{2\pi
r\sigma_{zz}}=\frac{\sqrt{3}q\hbar^2\nu}{4e^2\Gamma_0r}-
i\omega\frac{\sqrt{3}q\hbar^2}{4e^2\Gamma_0r}=R_0-i\omega L_0. \f
Here, $R_0$ and $L_0$ have meaning and magnitudes on the order of quantum resistance and kinetic inductance, respectively which are defined in the framework of the transmission line model for metallic CNTs \cite{BurkeNT,Maffucci}.
\subsection{Eigenwaves in metamaterials made of infinitely long CNTs and in finite-thickness slabs of carbon nanotubes}

\begin{figure}
\includegraphics[width=.9\columnwidth]{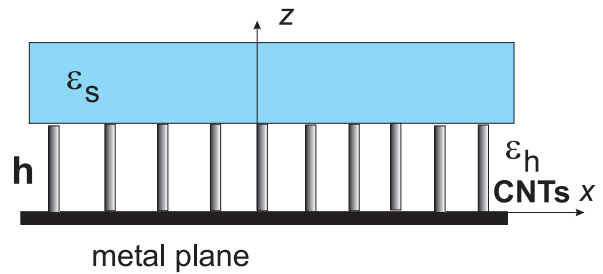}
\caption{Geometry of free-standing CNTs.} \label{geom}
\end{figure}

Finite-thickness slab of CNTs, embedded into a host matrix with the relative permittivity $\E_h$, is shown in Fig.~\ref{geom}. It forms two-dimensional periodic
structure in the $xy$-plane with the square lattice (for simplicity) and the lattice constant $d$.
The radius of the CNT $r$ and the lattice constant $d$ are taken to be $r=0.82$\,nm and $d=15$\,nm, respectively.
For the eigenwaves propagating in the arrays of infinitely long carbon nanotubes, space-time dependence of fields and currents is taken as $\exp{[-i(\omega t-k_xx-k_zz)]}$. The  approximation used in the following consideration is adopted from \cite{IgorPRB}.
The surface conductivity for thin CNTs is highly anisotropic; namely, the axial component of the conductivity tensor $\sigma_{zz}$ is much larger than its azimuthal component due to quantum effects \cite{Slepyan}. For this reason one can neglect by the azimuthal current on a CNT surface compared with the axial current. As a consequence, one can also neglect $z$-component of the magnetic field and as a result, Maxwell's equations are split into two subsystems describing the TM and TE waves.  The following consideration will relate to TM-modes, whose electric field lies in the $xz$-plane because fields of TE-modes do not interact with thin CNTs.
 The effective medium model \cite{Minsk,PNF} will be employed.
In the framework of the effective medium model, the CNT
array can be considered as a uniaxial material with the relative permittivity
dyadic
\e\l{p1} \=\E=\E_{zz}\_z_0\_z_0+\E_h(\_x_0\_x_0+\_y_0\_y_0).
\f
Transverse components of the diagonal permittivity tensor are equal to $\E_h$ since carbon nanotubes do not affect the electric field vector, orthogonal to them due to very small thickness and very low azimuthal conductivity of CNTs.
 As shown in \cite{Minsk},
\e\l{p2}\begin{array}{cc}
{\ds \frac{\E_{zz}}{\E_h}=1-\frac{k_p^2}{k^2+i\xi k}}, &
{\ds k_p^2=\frac{\mu_0}{d^2L_{\rm cnt}}},
\end{array}
 \f
where $k$ is the wavenumber in free space, $k_p$ is
the effective plasma wavenumber, and $L_{\rm cnt}$ is the effective inductance per unit length which includes the kinetic inductance $R_0$ and the electromagnetic inductance which can be neglected compared to the kinetic one, so $L_{\rm cnt}\simeq L_0$ \cite{BurkeNT}; $\mu_0$ is the permeability of vacuum and the parameter $\xi=\sqrt{\E_0\mu_0}(R_0/L_{\rm cnt})$, where $\E_0$ is the permittivity of vacuum, is responsible for losses.
Thus
the material behaves as a uniaxial free-electron plasma, where electrons can move only
along $z$-direction.
Apparently, this corresponds to an indefinite medium at frequencies below the
plasma frequency, because the permittivity in the directions
orthogonal to the tubes  equals to $\E_h$.
Dispersion equation for waves propagating in a uniaxial crystal is \cite{Land}

\e\l{y1}
\E_h k_{\perp}^2=\E_{zz}(k^2-k_z^2),\f
where $k_{\perp}^2=k_x^2+k_y^2$.
Substituting \r{p2} into Eq.~\r{y1} one obtains:
\e\l{p7}
k_z^2=k^2\E_h-k_\perp^2\dfrac{k^2+ik\xi}{k^2-k_p^2+ik\xi}.\f
It describes a typical conic-type dispersion. This cone is similar to the  surface derived for the numerical model which  accounts for periodicity of the CNT array (e.g., Fig.~3 from \cite{IgorPRB}).
Applicability of the effective medium model  was verified
by comparing with the results of the electrodynamic model \cite{IgorPRB}. An excellent agreement with
more accurate theory occurs within a wide range of transversal wave numbers, at least for $|k_{\perp}|<1000\,k$, in the mid-IR range for $d=15$\,nm \cite{Giant}.

\subsection{Waveguide properties of finite-thickness slabs of carbon nanotubes}

The following is to show that backward waves can propagate in the arrays of vertically standing CNTs. First, for simplicity consider CNTs placed between perfect electric conductor (PEC) and perfect magnetic
conductor (PMC) planes, where the PMC boundary models the open-ended interface with free space.
Assuming that waves propagate in the $x$-direction, i.e. $k_{\perp}=k_x$,
the relation between the transversal wave
vector component $k_x$ and the wavenumber in free space follows from Eq.~\r{p7}:
\e\l{p8}
k_x^2=\left[1-\dfrac{k_p^2}{k^2+ik\xi}\right](k^2\E_h-k_{z}^2),\f
where $k_z=m\pi/(2h)$, $m$ is the integer determining a number of field variations along CNTs (the transverse resonant condition).
Assuming $\xi=0$, one can readily show that the
derivative ${\rm d}k_{x}^2/{\rm d}k^2<0$, if $k_z/k>1$ and $k_p/k>1$.
 Thus, in this regime the finite-thickness slab supports propagation of \emph{backward waves}, because the group velocity is in the opposite direction to phase velocity.
 This property can be understood also from considering a planar waveguide
 filled with a CNT array (the array axis is orthogonal to the walls of the waveguide).  The propagation constant of the TM mode along  the waveguide is equal to
 \e\l{w1}
 k_{x}=\sqrt{(\E_{zz}/\E_h)\left[k^2-(m\pi/2h)^2\right]},\f
 where $m$ is a positive integer and $h$ is the height of the waveguide \cite{SpatD}.
If $\E_{zz}<0$, backward-wave propagation is allowed
 when $k<m\pi/2h$ and forbidden for $k>m\pi/2h$.

\begin{figure}
\includegraphics[width=.7\columnwidth]{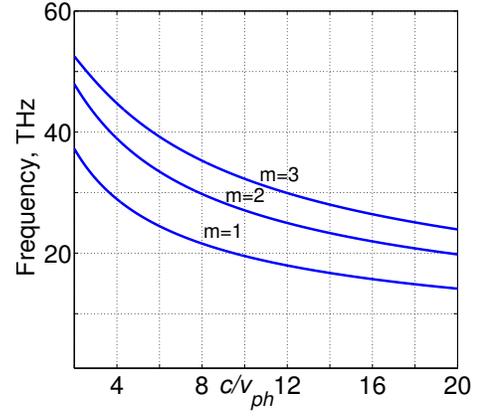}
\caption{Dispersion of three lowest modes.} \label{PEC_PMC}
\end{figure}

Figure~\ref{PEC_PMC} illustrates dispersion of this idealized structure. Here, all TM modes are backward waves;  $\omega~<~\omega_p~=~k_pc$; the plasma frequency $\omega_p/2\pi$  is calculated  to equal 58.7~THz.

Waveguide field components read as
\e\l{wg}\left.\begin{array}{l}
E_x(x,z)=A_m\sin{k_zz}\\
E_z(x,z)=A_m({k_x\E_h}/{k_z\E_{zz}})\cos{k_zz}\\
H_y(x,z)=-({A_m}/{Z})\cos{k_zz}
\end{array}\right\}
\exp{(ik_xx)}.
\f
where $A_m$ is the amplitude of the $m$-th mode and
\e\l{zi}
Z=-E_x/H_y=\eta k_z/k\E_h\f
is the transverse wave impedance for the TM waves. Here $\eta$ is the wave impedance of vacuum.
The spectrum of TE modes is the same as in a parallel-plane waveguide, filled with the dielectric with permittivity $\E_h$; CNTs do not affect the TE modes propagation. All TE modes are forward ones and are not interesting for our purposes.
Compared to optical MM made of resonant elements like split-ring resonators, dramatically increased bandwidth of backward-wave propagation is seen that gives  the ground to consider CNT arrays as a
 \emph{perfect backward-wave metamaterial}.

Then consider a more realistic structure, namely, CNTs with open ends,  adjoining with a half-space, filled in with a dielectric, characterized with the relative permittivity $\E_s$, as shown in Fig.~\ref{geom}.
In the framework of the effective medium theory the $2\times 2$ transfer matrix method \cite{Born}
can be used analyzing this structure.
The transfer matrix components $M_{mn}$ for the layer of carbon nanotubes have the form:
\e\l{tm}\begin{array}{ll}
M_{11}=M_{22}=\cos{k_{z}h}, & M_{12}=-iZ\sin{k_{z}h}, \\
M_{21}=-i(\sin{k_{z}h})/{Z}, &
\end{array}\f
where $Z$ is defined by formula \r{zi}.
The transfer matrix \r{tm} allows recalculation of tangential components of the electric and magnetic fields from the plane $z=0$ to $z=h$.
In contrast to the case of the plane plate waveguide bounded by PEC and PMC planes, $k_z$ is not fixed and  depends on the waveguide propagation constant $k_x$ according to Eq.~\r{p7}.
The transfer matrix method allows easy calculations for any problems related to multilayered structures. For the simple case of  surface waves propagating in the slab of CNTs with open ends,  whose fields attenuate exponentially from the interface with surrounding medium, which is the dielectric with permittivity $\E_s$ in our case, dispersion
$k(k_x)$ is given by the equation \cite{PNF}:
\e\l{dis}
({k_z}/{\E_h})\tan{(k_zh)}=\left({\sqrt{k_{x}^2-k^2\E_s}}\right)/{\E_s},\, k_x^2>k^2\E_s
\f
 which can be solved numerically.

\subsection{Eigenmodes, phase matching and losses}
\begin{figure}
\centering\includegraphics[width=.7\columnwidth]{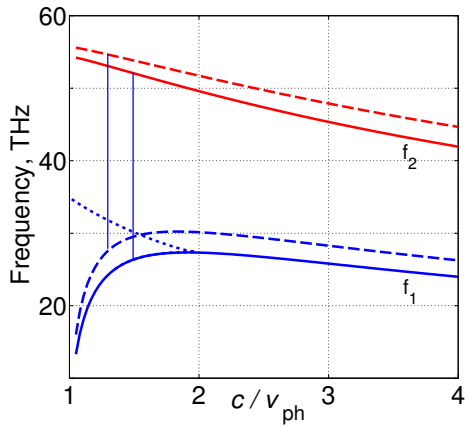}
\caption{Dispersion of two lowest eigenmodes in the slab of standing CNTs with open ends { and $\E_h=\E_s=1$}.
Real part of the normalized propagation constant $k_x'/k$ is shown for $h=1.05\, \mu$m (the solid lines), and $h=0.85\, \mu$m (the dashed line). Vertical lines mark  frequencies of the phase matched ordinary fundamental and backward second harmonic waves for thicknesses $h=1.05\, \mu$m and $h=0.85\, \mu$m. The dotted line shows  real part of  the normalized propagation constant for complex wave, existing in the stop band and calculated for $h=1.05\, \mu$m.} \label{open}
\end{figure}

Complex propagation constant $k_x=k_x'+ik_x''$ was calculated by numerically solving Eq.~\r{dis}.
Dispersion diagram for two lowest modes, calculated for two different thicknesses of the CNT layer, is shown in Fig.~\ref{open}.
Here, reduced wave vector $k_x'/k=c/v_{ph}=n_{ph}$ is a slow-wave factor (the ratio of speed of light $c$ to the phase velocity), which represents effective refraction index  $n_{ph}$.
Frequencies and dispersion of the allowed propagating EMW (eigenmodes) are determined by the
thickness and effective parameters of the metaslab and, hence, \emph{can be adjusted}. Appearance of the positive dispersion for small slow-wave factors and  stop-light regime
can be explained taking into account that directions of the Poynting vector in the metaslab and in the upper bounding dielectric (here, air) are opposite and the  total  energy flux stops if overall energy flows in both areas become equal.
 The  branch of $k_x'/k$ which descends down to  the stop light point of the dispersion curve at approximately  27.5~THz for $h=1.05\, \mu$m is characterized by a complex constant corresponding to a huge damping. (Such a branch is not shown for $h=0.85\, \mu$m.) Imaginary part,   $k_z''/k$, is not shown.
Waves with complex propagation constants appear in a stop band  if dispersion changes a sign and exist even in a lossless case (see \cite{radiofiz}, also \cite{DWM}, Fig.~6).
These waves do not transfer energy because  can be excited only as pairs of complex conjugate solutions and their fields can be represented in the form of standing waves with decaying amplitudes [see \cite{DWM}, Eq.~(28)].

 Phase of  NLO polarization oscillation at 2$\omega_1$ at a given point in the metamaterial is determined by its wave vector 2$\mathbf{k}(\omega_1)$, whereas that of the emitted SH -- by the wave vector $\mathbf{k}(2\omega_1)$. Hence, to achieve phase matching, the phase velocities of both waves, $v_{1,\rm ph}=\omega_1/k(\omega_1)$ and $v_{\rm SH, ph}=2\omega_1/k(2\omega_1)$ must be equal. Such a  possibility  for ordinary fundamental and BWSH waves  is shown in Fig.~\ref{open}. Here, the vertical lines show frequencies $f_2=2f_1$  and $f_1$ that correspond to one and the same slow-wave factors $c/v_{\rm ph}$.
For $h=1.05\, \mu$m,  $f_2$= 52.37 THz and $c/v_{\rm ph}\approx 1.5$, whereas for $h=0.85\, \mu$m,  $f_2$= 55 THz and $c/v_{\rm ph}\approx 1.27$.
  For $h=1.05\, \mu$m, the wavelength of the the wave at $f_1$  in CNTs is about $9\,\mu$m,  which corresponds to about $12\,\mu$m in free space.  Figure~\ref{open} demonstrates that positions and dispersion of  TM modes depend on the length of CNTs. It proves the possibility to ensure and to \emph{adjust} to different frequencies the co-existence of ordinary fundamental and BWSH  waves with \emph{equal} phase velocities and \emph{contra-directed} energy fluxes.

\begin{figure}[!h] 
\centering
\includegraphics[width=.7\columnwidth]{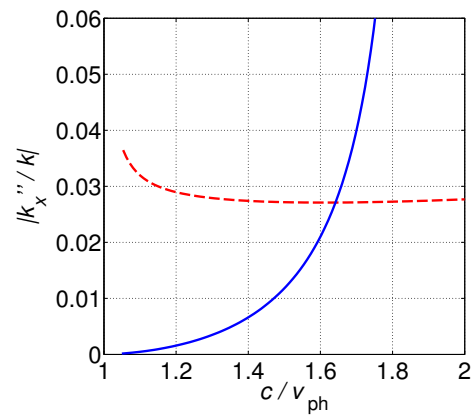}
\caption{Normalized $k_x''$ for the first mode (the blue solid plot) and for the second mode (the dashed red plot) at $h=1.05\, \mu$m.}
 \label{beta2}
 \end{figure}

Optical losses of EMWs  are represented by imaginary part of the propagation constant $k_x=k_x'+ik_x''$.
 For carbon nanotubes, magnitude of  $k_x''$ is determined by the relaxation time $\tau$ at  frequencies below optical transitions. At low frequencies, including those close to a few
terahertz, $\tau$ at room temperature  is estimated as $3\times 10^{-12}$\,s
\cite{Slepyan,Hanson}. Fig.~\ref{beta2} shows an attenuation of both modes in the proximities of  frequencies of the phase matched fundamental and BWSH waves.
\begin{figure}[h!]
\includegraphics[width=.7\columnwidth]{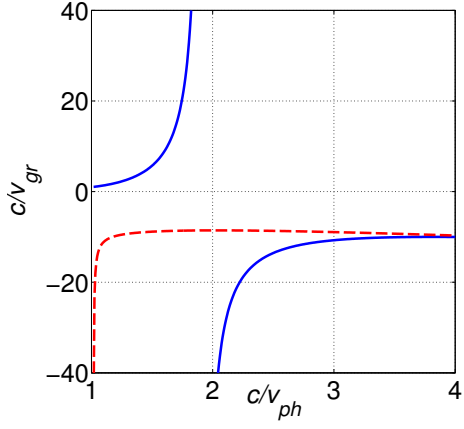}
\caption{Group velocity (the group delay factor) versus phase velocity (the slow-wave factor) for the same modes as in Fig.~\ref{open} at $h=1.05\, \mu$m. The red curve corresponds to the higher-frequency mode, the blue curve to the lower-frequency mode.}\label{group}
\end{figure}
In our one-dimensional case, Eq.~\ref{gr} reads as $v_{\rm g}=\partial\omega/\partial k_x$. The group delay factor $n_{\rm gr}=c/v_g$  for both modes is shown in Fig.~\ref{group}. It is seen that at $c/v_{\rm ph1,2}=1.5$, which corresponds to phase matching, $n_{\rm g,1}\approx 5.5$ and   $n_{\rm g,2}\approx 8.9$. The group delay factor is related with the density of modes (DOM) \cite{Aguanno}, which can be defined in our case as $\rho_{\omega}=dk_x/d\omega=n_{\rm g}/c$.

Note, that carbon nanotubes themselves possess nonlinear properties which can be used for harmonic generation \cite{PRA99,Guo,SlepHG,Morris,NemilenHG}.

\section{Backward-wave second harmonic generation in  pulsed regime}
In this section, SHG will be investigated for the case of fundamental EMW falling in the positive dispersion frequency domain and its SH in the negative one (Fig.~\ref{open}). To achieve phase matching, phase velocities of the fundamental and SH waves must be co-directed. This means that the pulse of SH  will propagate \emph{against} the pulse of the fundamental radiation. Such extraordinary NLO coupling scheme dictates {exotic} properties of the \emph{ frequency-doubling  ultracompact
NLO mirror} under consideration \cite{Mir,APA2016}.
Our goal is to investigate  the characteristic magnitudes of the metamaterial nonlinearity and of the fundamental field intensity that would enable an appreciable output of BWSHG  provided that the phase-matching is achieved. A simplified model of plane travelling waves will be used for such a purpose.
\subsubsection{Basic Equations}\label{pul}
Electric and magnetic components of the waves and corresponding nonlinear polarizations at $\omega_1$ and  $\omega_2~=~2\omega_1$ are defined as
\begin{eqnarray}
\{{\mathcal{E}}, {\mathcal{H}}\}_j=\Re{\{{E},{H}\}_j\exp\{i(k_jx-\omega_jt)\}},\label{eh}&&\\
\{\mathcal{P}, \mathcal{M}\}^{NL}_j=\Re{\{P, M\}^{NL}_j\exp\{i(\widetilde{k}_jx-\omega_jt)\}},\label{pme}&&\\
\{P, M\}^{NL}_1=\chi^{(2)}_{e,m,1}\{E, H\}_2\{E, H\}_1^*,\label{pm1}&&\\
 \{P, M\}^{NL}_2=\chi^{(2)}_{e,m,2}\{E, H\}_1^2,\, 2\chi^{(2)}_{e,m,2}=\chi^{(2)}_{e,m,1}. \label{pm2}&&
\end{eqnarray}
Amplitude $E_1$ of the first harmonic (FH) and of the SH, $E_2$,  are given by the equations:
\begin{eqnarray}
 s_2\frac{\partial E_2}{\partial x}+ \frac1{v_2}\frac{\partial E_2}{\partial t}=\nonumber&&\\ 
 =- i\frac{k_2}{\epsilon_2}4\pi\chi^{(2)}_{e,2}E_1^2\exp{(-i\Delta kx)}-\frac{\alpha_2}2E_2,\label{eqE2}&&\\
  s_1\frac{\partial E_1}{\partial x}+ \frac1{v_1}\frac{\partial E_1}{\partial t}=\nonumber&&\\ 
=  - i\frac{k_1}{\epsilon_1}4\pi\chi^{(2)*}_{e,1}E_1^*E_2\exp{(i\Delta kx)}-\frac{\alpha_1}2E_1.\label{eqE1}&&
 \end{eqnarray}
Here,  $v_i>0$ and $\alpha_{1,2}$  are  group velocities and absorption indices at the corresponding frequencies,  $\Delta k=k_{2}-2k_{1}$.   Parameter $s_j=1$ for ordinary, and $s_j=-1$ for backward wave.
With account for $k^2=n^2(\omega/c)^2$, $n_1=s_1\sqrt{\epsilon_1\mu_1}$, $n_2=s_2\sqrt{\epsilon_2\mu_2}$, we introduce effective nonlinear susceptibility $\chi^{(2)}_{\rm eff}=\chi^{(2)}_{e,2}=\chi^{(2)}_{e,1}/2$,
 amplitudes $e_{j}=\sqrt{|\epsilon_j|/k_j}E_j$ and  $a_j=e_i/e_{10}$, the coupling parameters $\ka=\sqrt{k_1k_2/|\epsilon_1\epsilon_2|} 4\pi\chi^{(2)}_{\rm eff}$ and $g=\ka E_{10}$, the loss  and phase mismatch parameters $\widetilde \alpha_{1,2}=a_{1,2}L$ and  $\Delta \widetilde{k}=\Delta {k}l$, the metaslub thickness $d=L/l$, the position $\xi=x/l$ and the time instant $\tau=t/\Delta\tau$. It is assumed that   $E_{j0}=E_j(x=0)$, $l=v_1\Delta\tau$ is the pump pulse length and $\Delta\tau$ is duration of the input fundamental pulse. Quantities  $|a_j|^2$ are proportional to the time dependent photon fluxes. Then Eqs. \eqref{eqE2} and \eqref{eqE1} are written as
\begin{eqnarray}
 s_2\frac{\partial a_2}{\partial \xi}+ \frac{v_1}{v_2}\frac{\partial a_2}{\partial \tau}=
  -igla_1^2\exp{(-i\Delta \widetilde{k}\xi)}-\frac{\widetilde{\alpha}_2}{2d}a_2, \label{eq2a}&&\\
    s_1\frac{\partial a_1}{\partial \xi}+\frac{\partial a_1}{\partial \tau}=
     -i2g^*la_1^*a_2\exp{(i\Delta \widetilde{k}\xi)}-\frac{\widetilde{\alpha}_1}{2d}a_1. \label{eq2b}&&
\end{eqnarray}

In the case of magnetic nonlinearity, $\chi^{(2)}_{\rm eff}=\chi^{(2)}_{m,2}$, equations for amplitudes  $H_j$ take the form:
\begin{eqnarray}
 s_2\frac{\partial H_2}{\partial x}+ \frac1{v_2}\frac{\partial H_2}{\partial t}=\nonumber&&\\
 =- i\frac{k_2}{\mu_2}4\pi\chi^{(2)}_{m,2}H_1^2\exp{(-i\Delta kx)}-\frac{\alpha_2}2H_2,\label{eq1a}&&\\
  s_1\frac{\partial H_1}{\partial x}+ \frac1{v_1}\frac{\partial H_1}{\partial t}=\nonumber&&\\
 = - i\frac{k_1}{\mu_1}4\pi\chi^{(2)*}_{m,1}H_1^*H_2\exp{(i\Delta kx)}-\frac{\alpha_1}2H_1.\label{eq1b}&&
 \end{eqnarray}
Then, equations for amplitudes
  $a_j=m_i/m_{10}$, where $m_{j}=\sqrt{|\mu_j|/k_j}H_j$, take the form of \eqref{eq2a} and \eqref{eq2b}, with coupling parameters given by $\ka=\sqrt{k_1k_2/|\mu_1\mu_2|} 4\pi\chi^{(2)}_{\rm eff}$ and $g=\ka  H_{10}$. In both cases, $x_0=g^{-1}$ is characteristic medium length required for significant NLO energy conversion.
Only the most favorable case of exact phase matching $\Delta {k}=0$  will be considered below.

The following values and estimates, which are relevant to the MM made of nanotubes of height  $h=1.05$$\mu$m (Figs.~\ref{beta2} and \ref{group}), were used at numerical simulations. For  pulse duration of
$\Delta\tau=10$~ps, the spectrum bandwidth  is on the order of $\Delta f\approx 1/\Delta\tau=0.1$~THz. Hence,  $\Delta f/f \propto10^{-2}\div10^{-3}$, and  phase matching can be achieved for the whole frequency band. This becomes impossible at $\Delta\tau=10$~fs because $\Delta f/f \propto10$.
Phase matching is achieved at $k_1=2\pi f_1/c=5.47\times10^5$~m$^{-1}$, $k_2=10.95\times10^5$~m$^{-1}$ (Fig. \ref{open}). Then one finds $\alpha_1=2k_1^{''}=2\times9.3\times10^{-3}\times k_1=1.02\times10^4$~m$^{-1}$, $\alpha_2=2k_2^{''}=2\times2.72\times10^{-2}\times k_2=5.96\times10^4$~m$^{-1}$.
Since losses for the second mode is greater, the characteristic metaslab thickness corresponding to extinction $I/I_0=\exp(-\alpha_2L)=0.1$ is estimated as  $\alpha_2L=2.4$, $\alpha_1L=0.41$, $L\approx40\ \mu$m, which is about 10 wavelengths in the medium. The fundamental pulse  length is $l=\Delta\tau v_1=\Delta\tau c/n_{g,1}=10^{-11}\times6.06\times10^7\ m=606\ \mu$m, which is  15 times greater than $L$. The later indicates that whereas some transients occur at the pulse forefront and the tail, the quasistationary process establishes through almost the whole pulse duration. At $\Delta\tau\leq10$~ps, which is still acceptable, the effect of the transient processes significantly increases.

The input pulse shape was chosen close to a rectangular form
\begin{equation}
F(\tau)=0.5\left(\tanh\frac{\tau_0+1-\tau}{\delta\tau}-\tanh\frac{\tau_0-\tau}{\delta\tau}\right),
\end{equation}
where $\tau=t/\Delta\tau$, $\delta\tau$ is  duration of the pulse front and tail, and $\tau_0$ is  shift of the front relative to $t=0$.
Magnitudes $\delta\tau=0.01$ and $\tau_0=0.5$ were selected for numerical simulations. Coupling parameter $gl=l/x_0$ is proportional to strength of the input pump field and presents a ratio of the input pulse length and characteristic nonlinear conversion length.
\subsubsection{Transparent NIM slab}\label{pult}
 In order to show unparallel properties of the BWSHG in a frequency double domain positive/negative spatially dispersive slab, first, consider a loss-free slab that supports ordinary EMW at fundamental frequency and BEMW at SH frequency ($s_1=-1$, $s_2=1$).
\begin{figure}[h!]
\begin{center}
\includegraphics[width=.49\columnwidth]{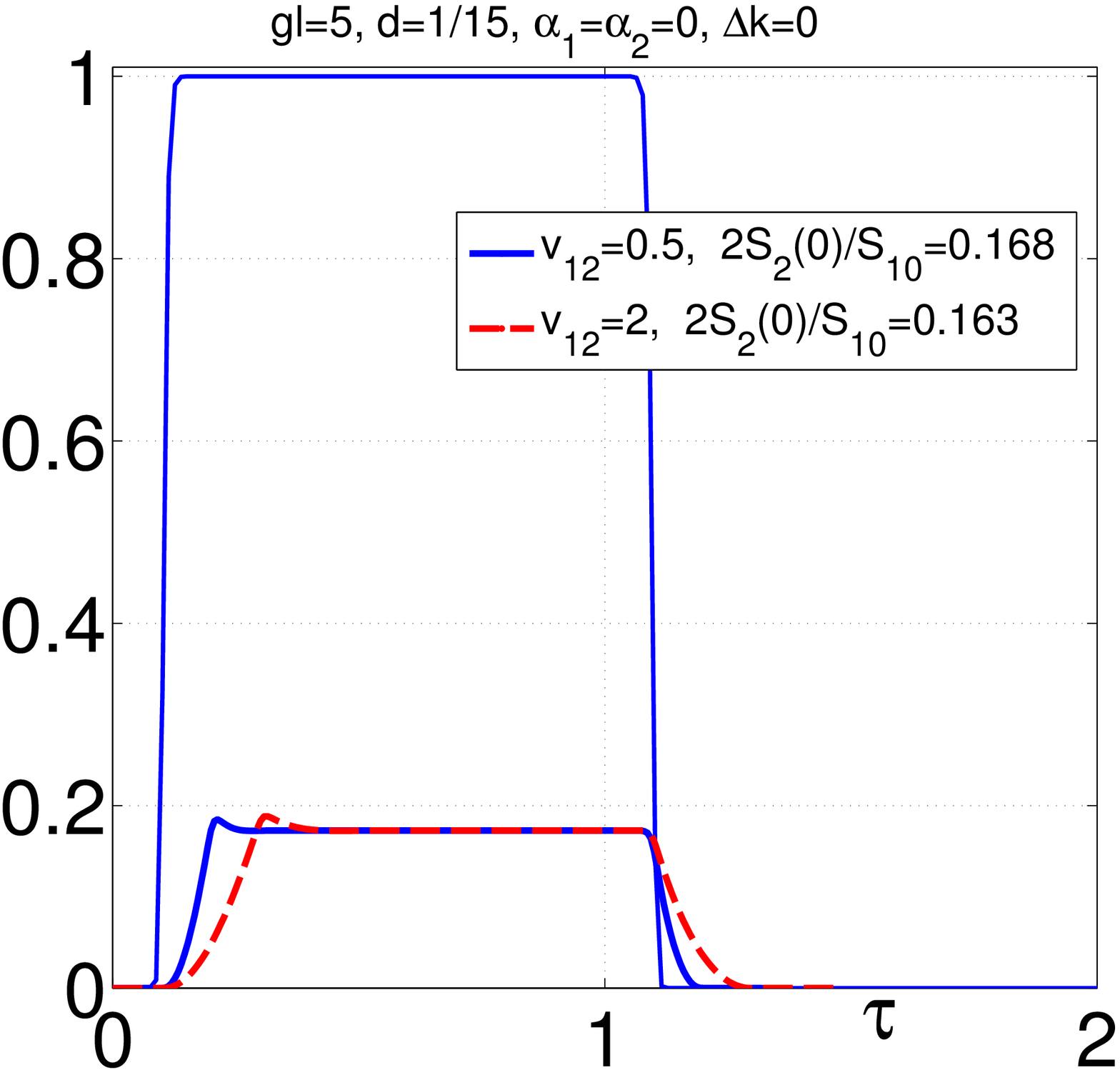}
\includegraphics[width=.49\columnwidth]{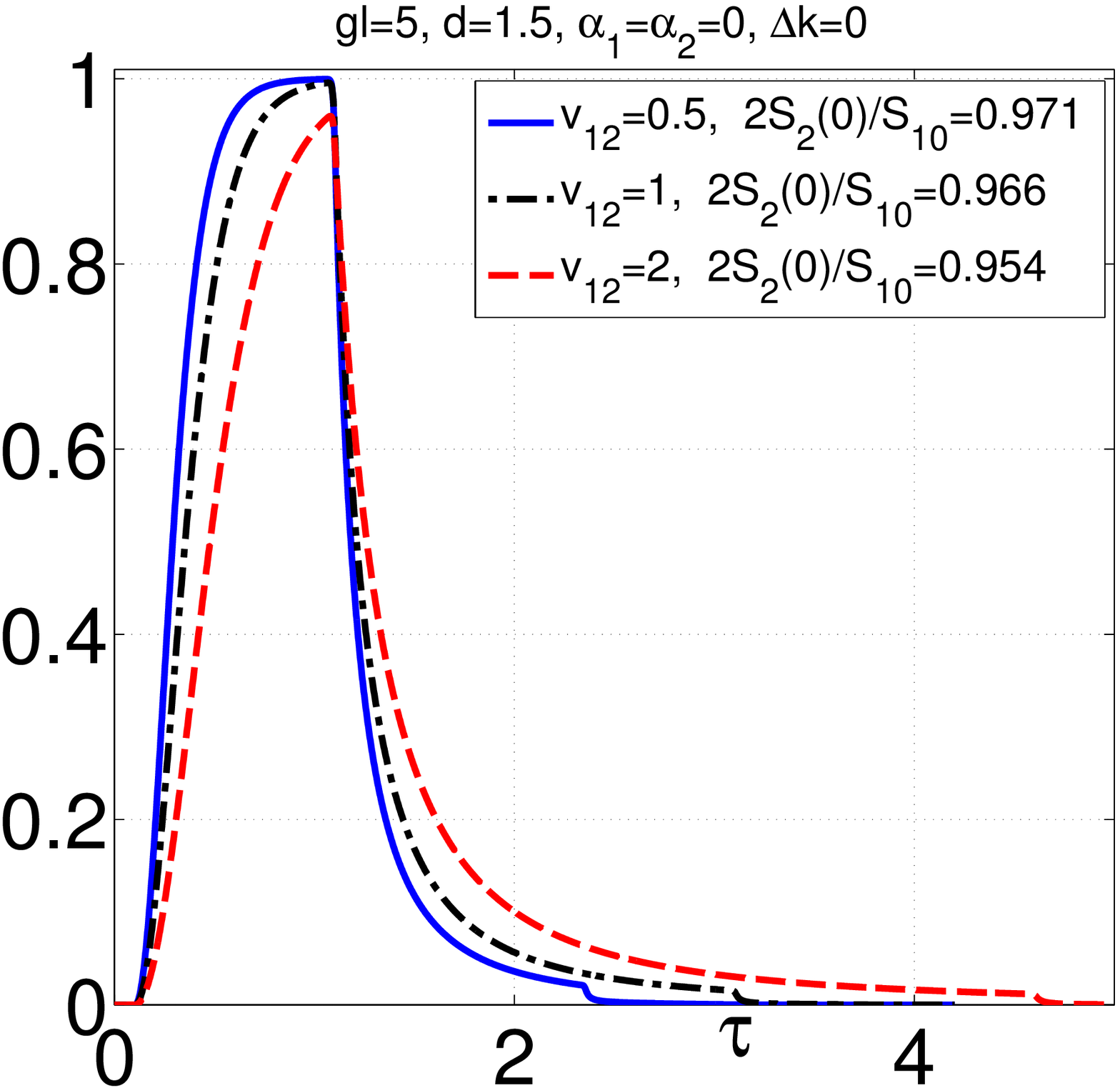}\\
(a)\hspace{0.5\columnwidth} (b)\\
\includegraphics[width=.48\columnwidth]{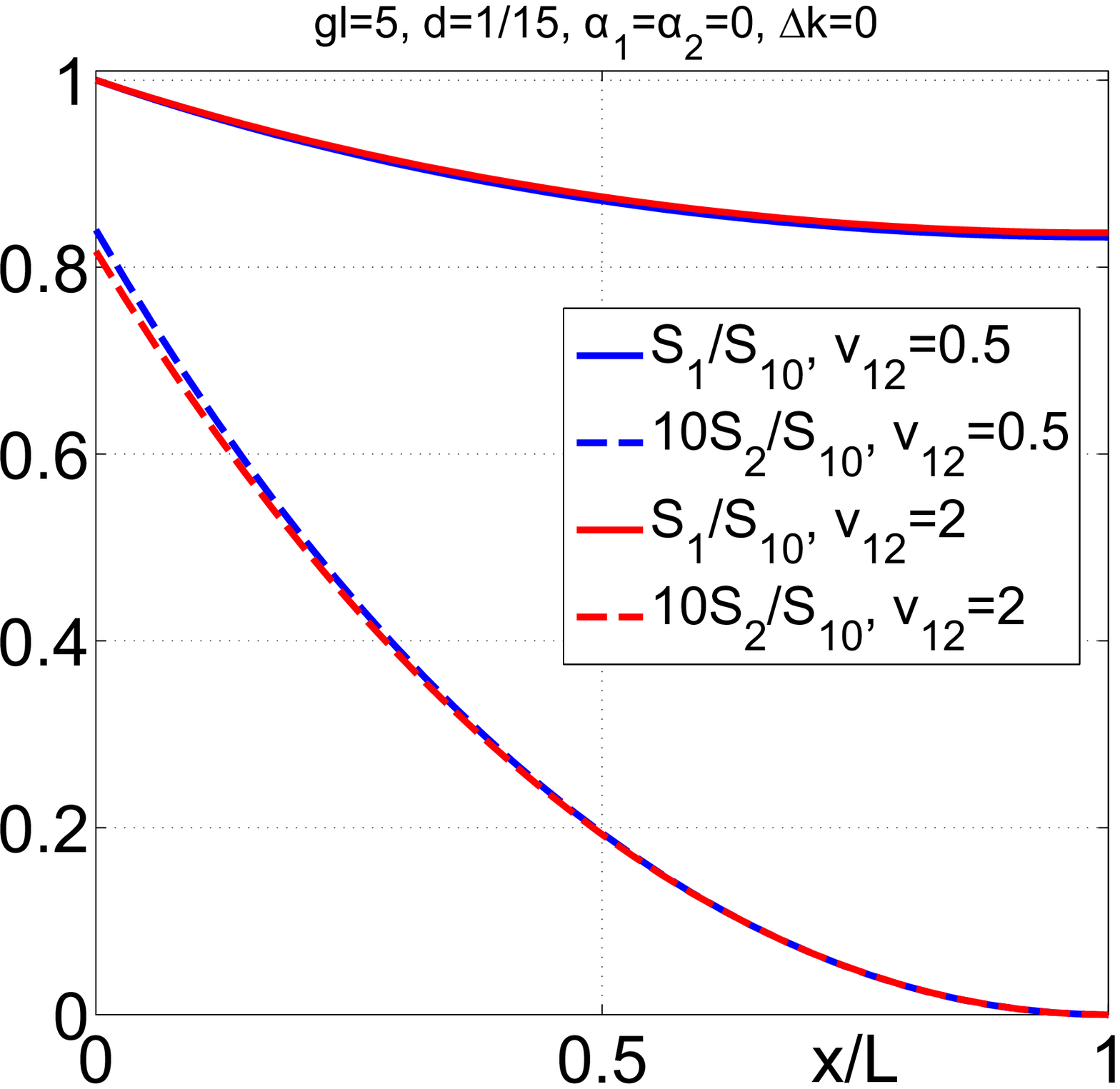}
\includegraphics[width=.49\columnwidth]{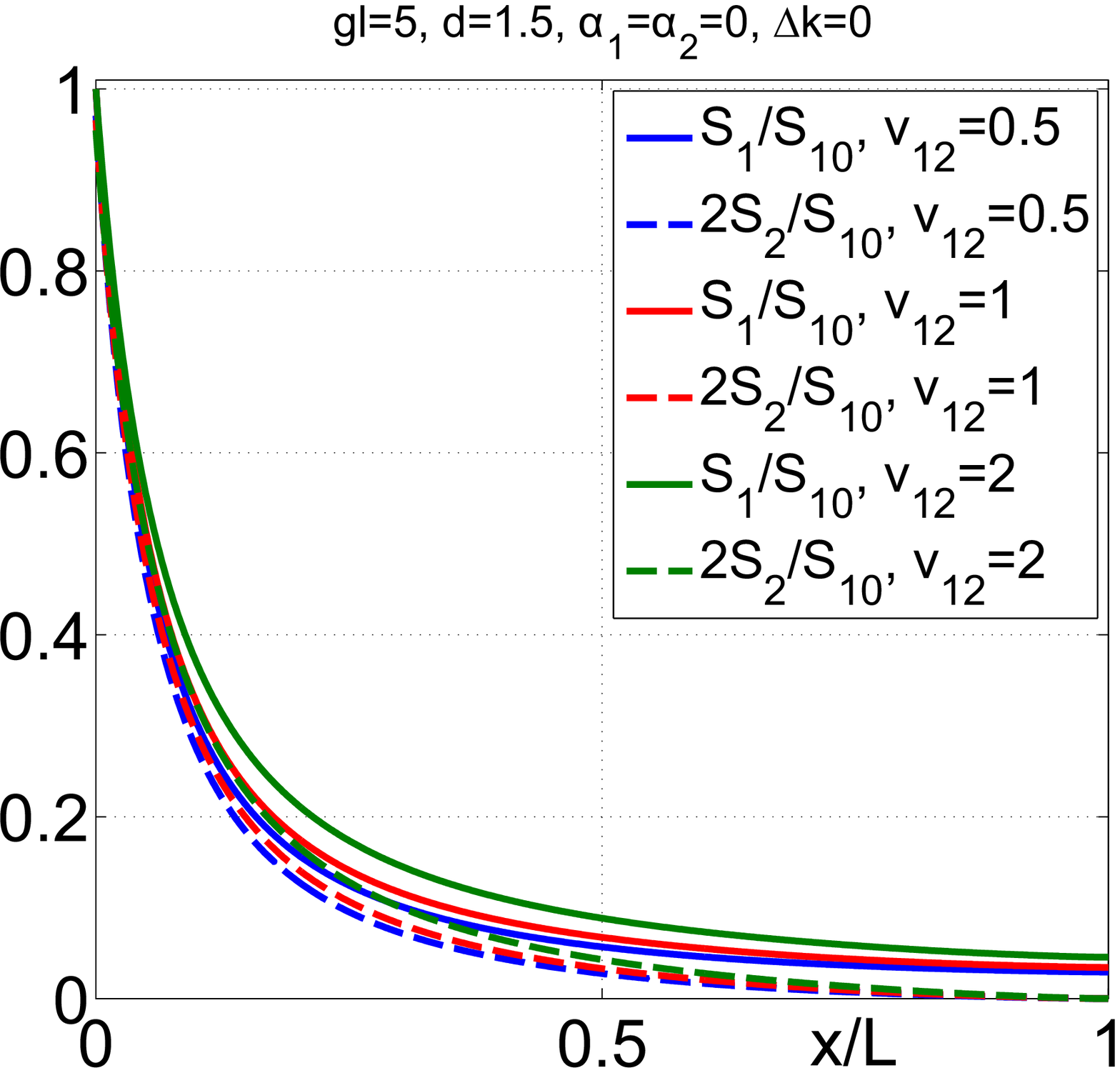}
(c)\hspace{0.5\columnwidth} (d)
\caption{\label{fi1} Backward-wave SHG in a transparent MM: dependence of shape and energy of output  pulses on the metaslab thickness and  group velocity dispersion $v_{12}=|v_1/v_2|$. $gl=5$. a)  Input fundamental pulse  at $x=0$ (the rectangular blue plot), output contra-propagating pulse of BWSH at $x=0$ (the red plot) and output fundamental pulse at $x=L$ (the lower blue plot) for two different ratios of group velocities at $L/l=1/15$.  b) Shape of output pulses of BWSH  at $L/l=1.5$ for different ratios of group velocities.  c) and d) Depletion of energy of the fundamental pulse along the metaslab and growth of energy of the BWSH pulse in the opposite direction for the same parameters as  in panels a) and b).}
\end{center}
\end{figure}
\begin{figure}[h!tp]
\centering
\includegraphics[width=.49\columnwidth]{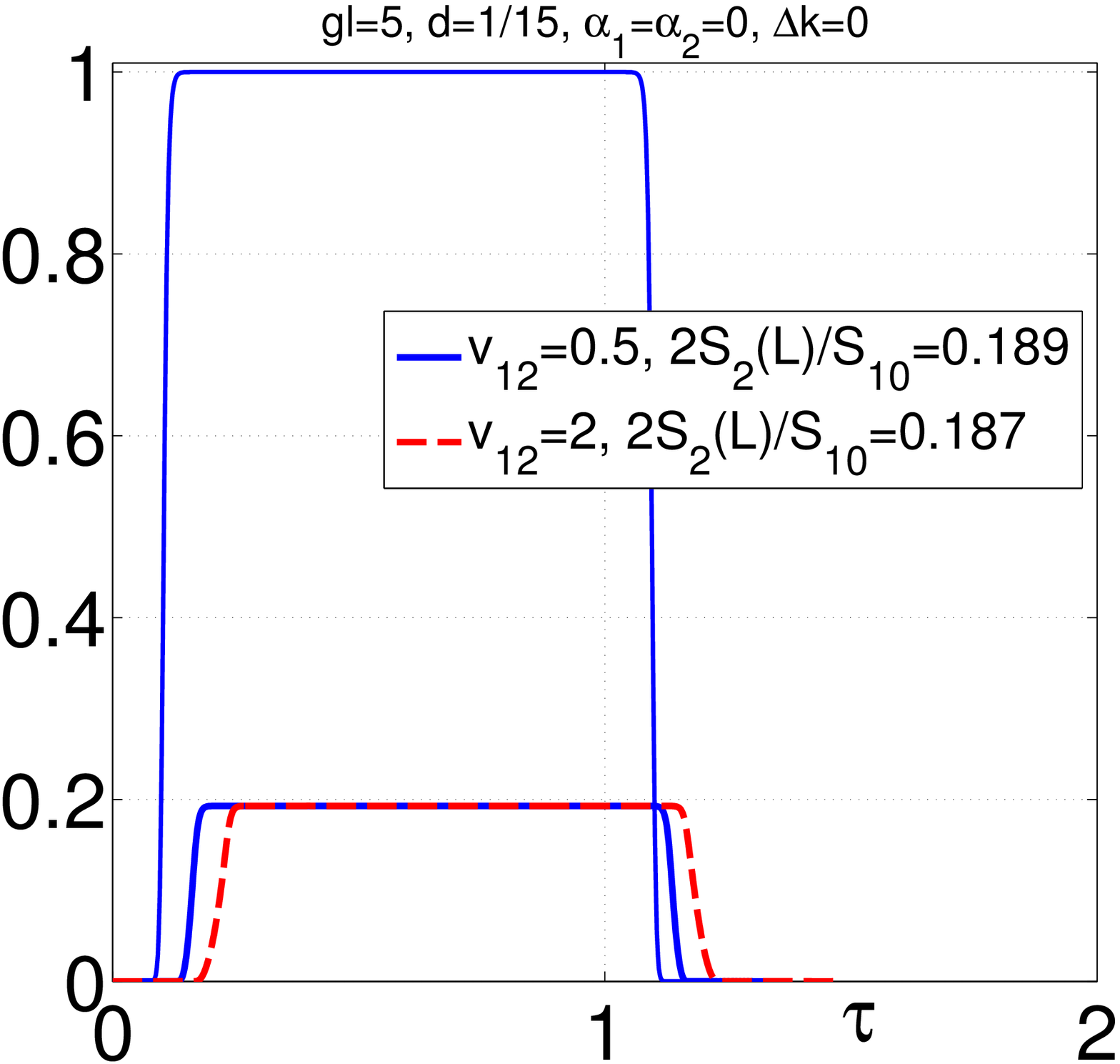}
\includegraphics[width=.49\columnwidth]{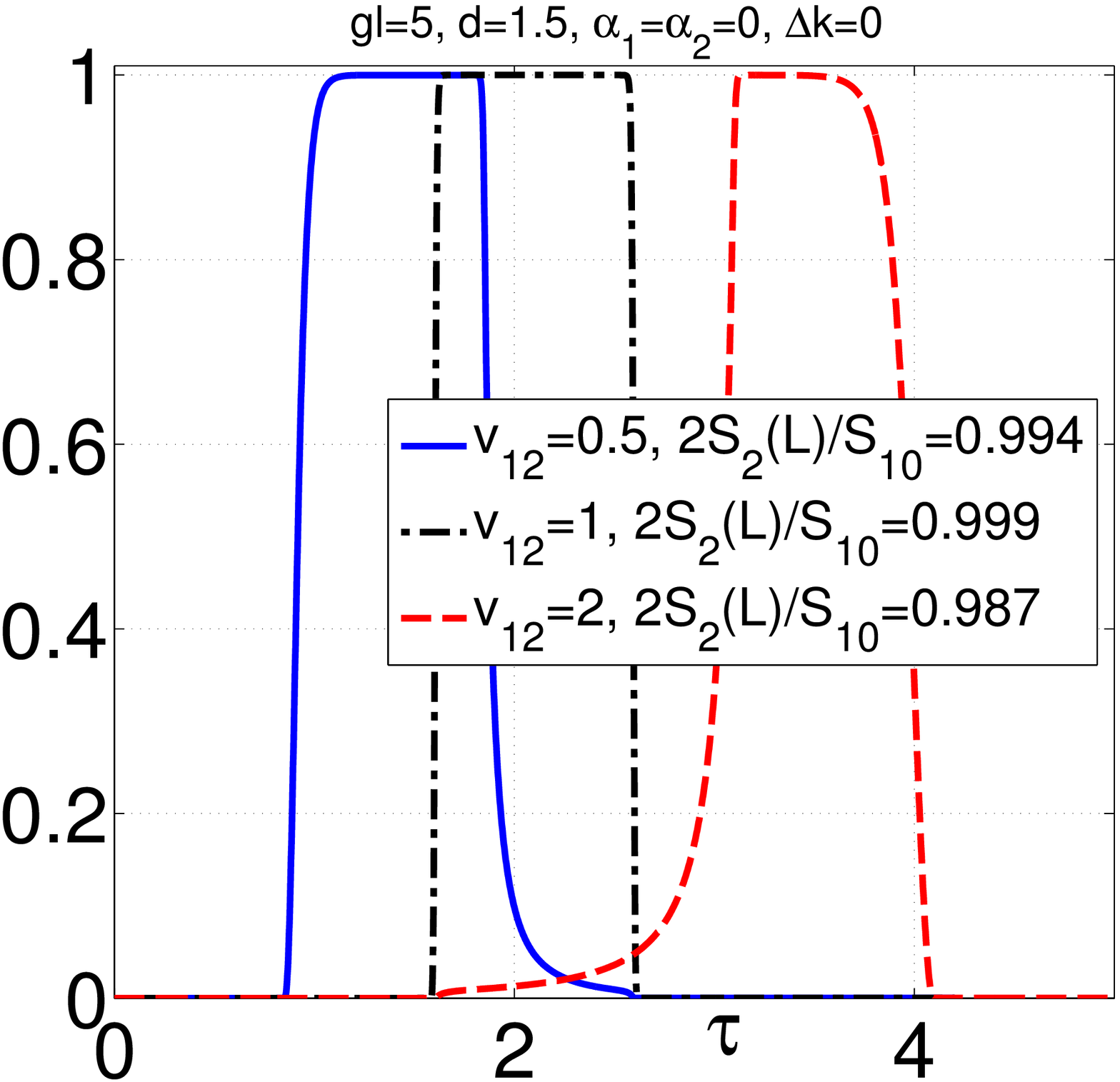}\\
\centering (a)\hspace{0.5\columnwidth}(b)\\
\centering
\includegraphics[width=.49\columnwidth]{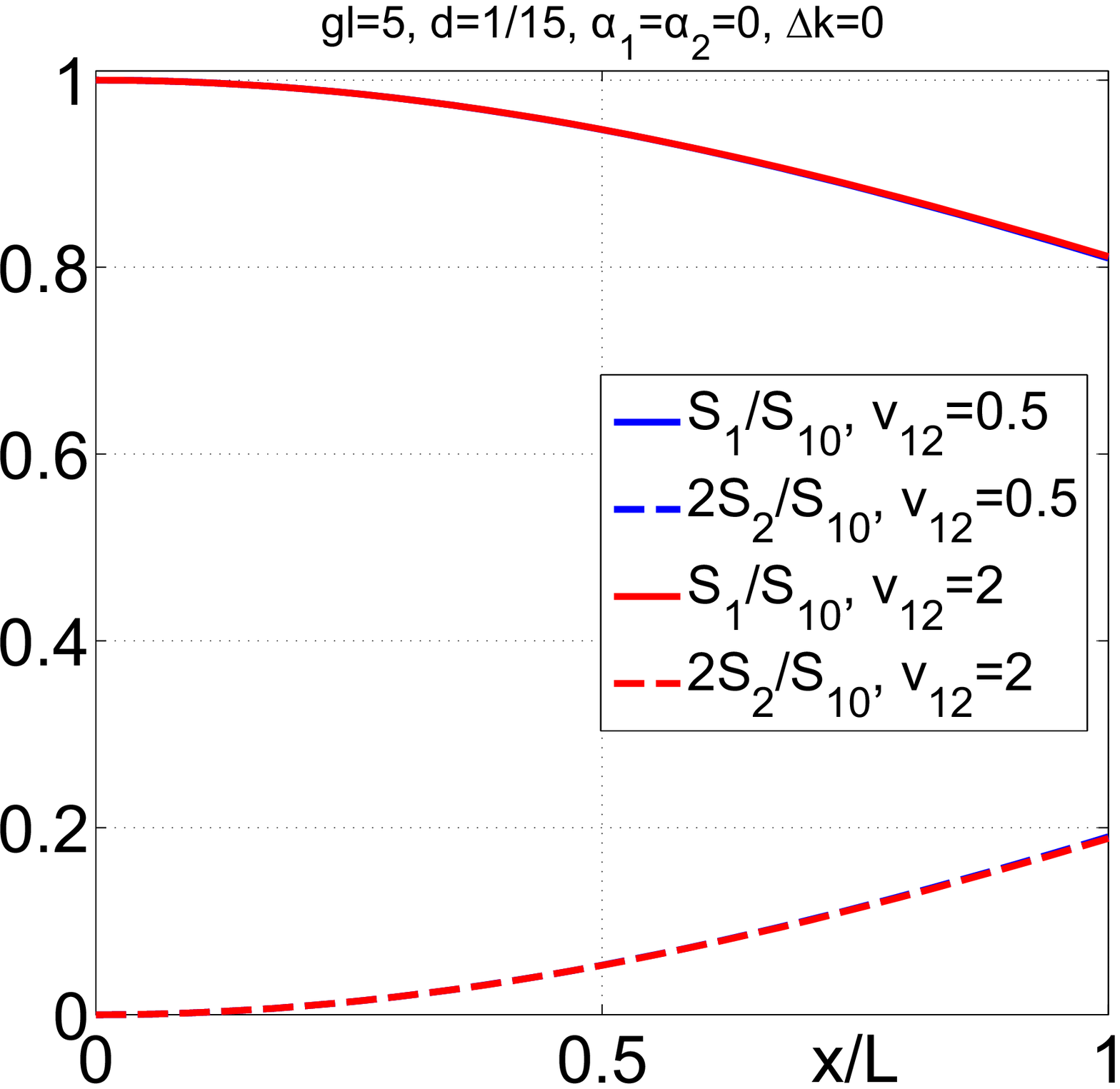}
\includegraphics[width=.49\columnwidth]{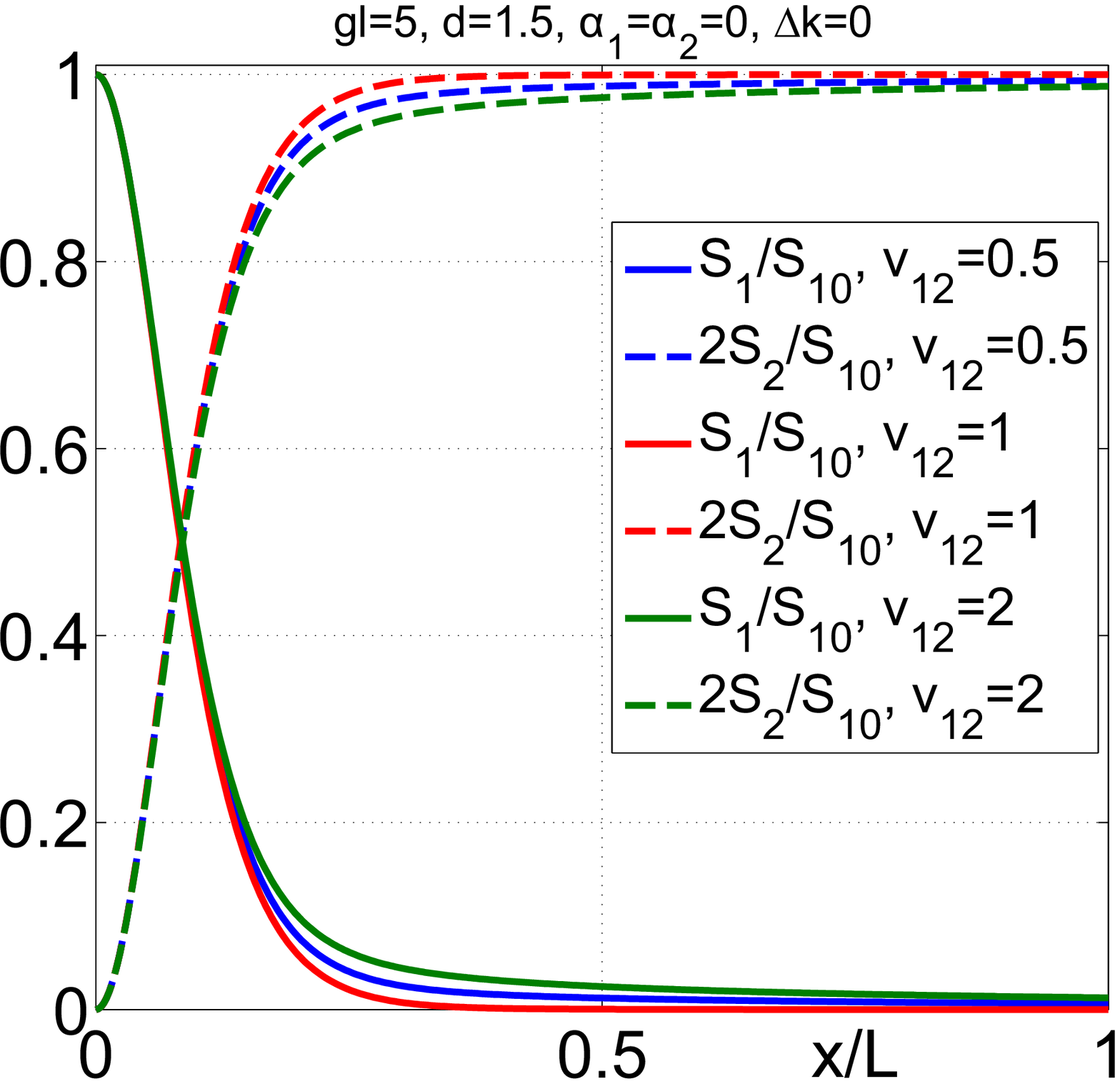}\\
\centering (c)\hspace{0.5\columnwidth} (d)
\caption{Ordinary SHG in a transparent material, all other parameters are the same as in Fig. \ref{fi1}.  } \label{trpim}
\end{figure}
Unusual properties of SHG in NIMs in the pulsed regime stem from the fact that it occurs only inside the traveling pulse of fundamental radiation. Generation begins on the leading edge of the pulse, grows towards its trailing edge, and then exits the pulse with no further changes. Since the fundamental pulse propagates across the slab, duration of the contra-propagating SH pulse appears longer than the fundamental one. Basically, depletion of the fundamental radiation along the pulse  and the conversion efficiency depend on its initial maximum intensity, phase  matching and group velocities of pulses. Ultimately,  properties of BWSHG are anticipated to depend on ratio of the fundamental pulse and the slab lengths.
Such extraordinary behavior is illustrated in Fig.~\ref{fi1}.
Figure~\ref{fi1}(a) displays  rectangular shape of the input fundamental pulse $T_1=|a_1(\tau)|^2/|a_{10}|^2$ entering the slab at $x=0$  (the blue rectangular plot). Shapes of the output pulses of BWSH traveling in  the opposite, reflection direction $\eta_2(\tau)=|a_2(\tau)|^2/|a_{10}|^2$ and exiting the slab at $x=0$  are depicted by the lower blue and red plots for the input pulse length $l$ which is 15 times longer than the metaslab thickness ($d=1/15$), and for two different group velocities.  This case is close to continuous-wave regime. Figure~\ref{fi1}(b) also show  results of numerical simulations  for shapes  of  output BWSH pulses for the same input pulse amplitude corresponding to $gl=5$,  however,  for the pulse length 1.5 times shorter than the metaslab thickness ($d=1.5$).  The plots demonstrate that output BWSH pulse becomes longer and its shape significantly differs from  the shape of  input fundamental pulse. Such a behavior is due to the fact that  non-stationary transient processes at pulse edges that does not play significant role in the first case become important in the second case.
Figures~\ref{fi1}(c) and (d) depict  the pulse energy (quantum) conversion efficiency  $\eta_2=S_2/S_{10}=\int dt |a_2(x)|^2/|a_{10}|^2$ and depletion of energy of the of the fundamental pulse $S_1(x)/S_{10}$ along the slab and at the corresponding exits for the same coupling parameter  $gl=5$.

Fundamental differences between BWSHG and SHG in ordinary materials are seen from comparison of Fig.~\ref{fi1} and Fig.~\ref{trpim}. In the first case, SH propagate against the fundamental pulse towards its greater magnitudes. The gap between the plots is inverse proportional to the conversion efficiency. It is seen from comparison of plots in Figs.~\ref{fi1} c and d that  energy conversion efficiency grows with increase of the ratio $d$ (thicker MM slabs) and is almost independent on the dispersion of the group velocities for the given range of the select parameters. The plots satisfy to  Manley-Rove (the energy conservation) law: the \emph{difference} between the the number of pairs of photons in FH and the number of photons in SH ($S_1/2-S_2$) is a constant along the slab, which magnitude decreases with increase of the parameter $gl$ \cite{Sl,APB,OL}. Total number  of annihilated pair of photons in the FH ($S_{10}-S_{1L})/2$ is equal to the number of output SH photons $S_{20}$. In contrast, for ordinary materials and co-propagating waves, a \emph{sum} of the the number of pairs of photons in FH and the number of photons in SH ($S_1/2+S_2$) is a constant along the slab. Shape, width and delay of the output pulses of SH generated in ordinary and BW materials is also differ.

\begin{figure}[htbp]
\begin{center}
\includegraphics[width=.49\columnwidth]{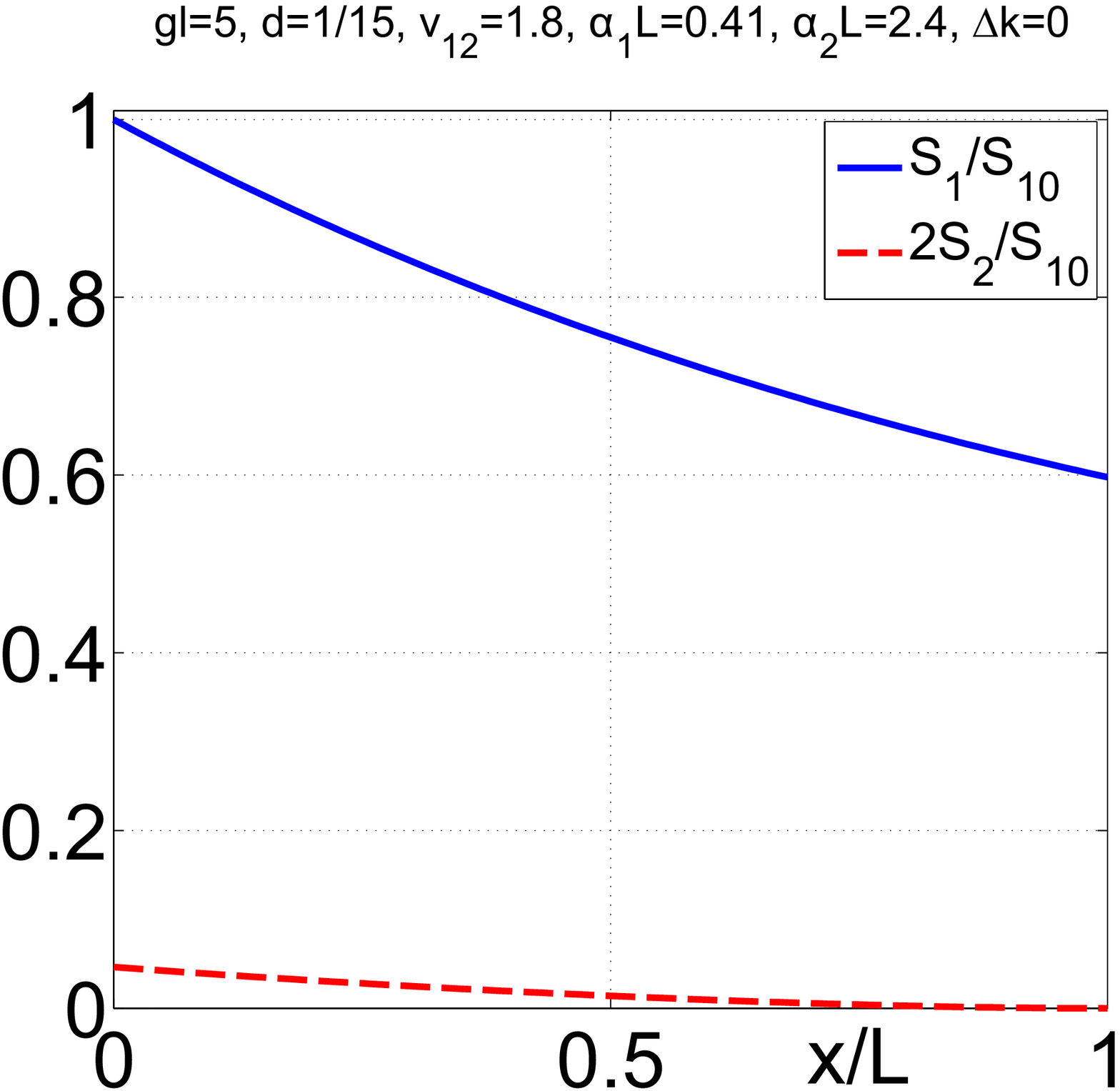}
\includegraphics[width=.49\columnwidth]{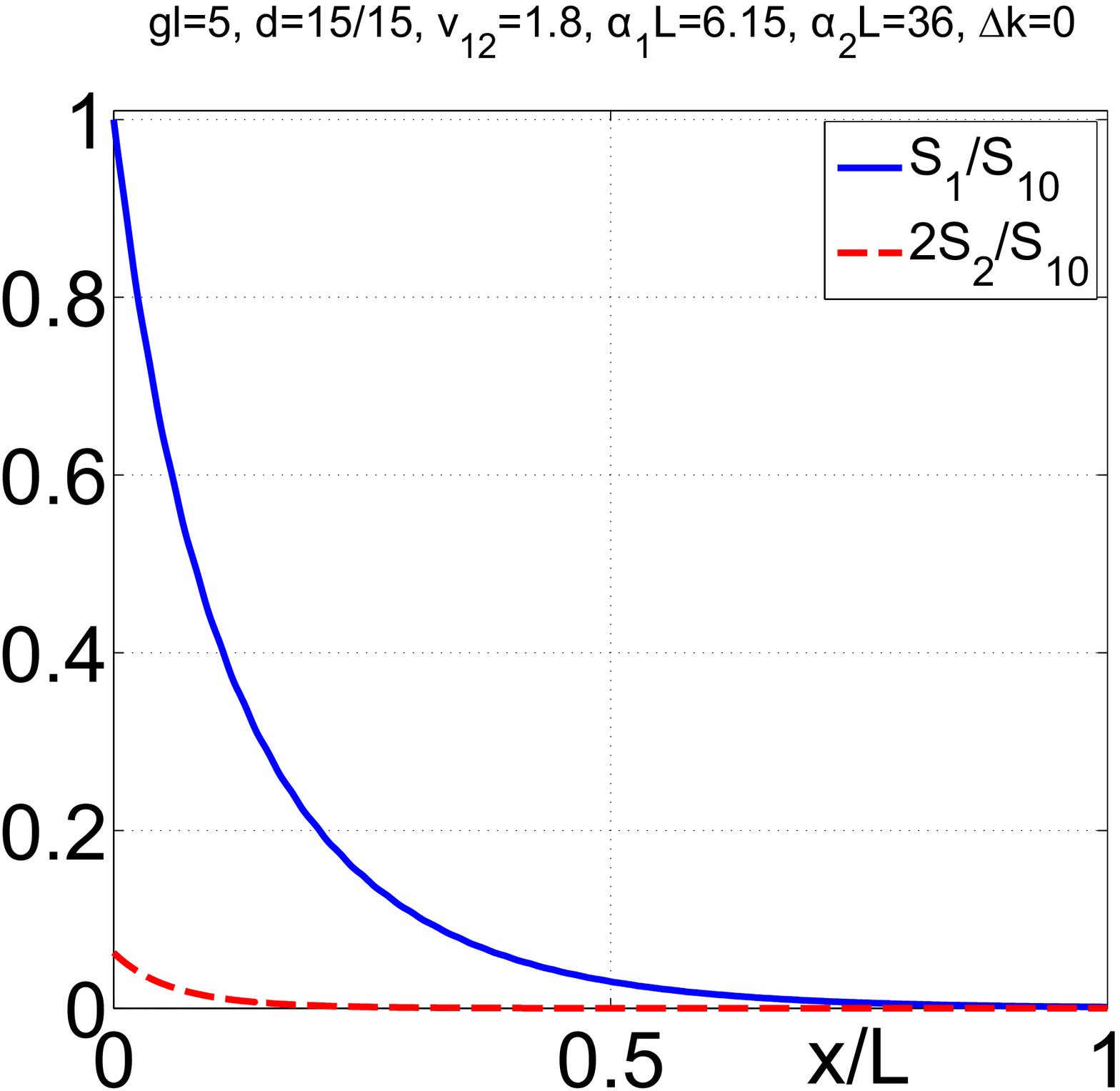}\\
 (a)\hspace{0.5\columnwidth} (b)\\
\includegraphics[width=.49\columnwidth]{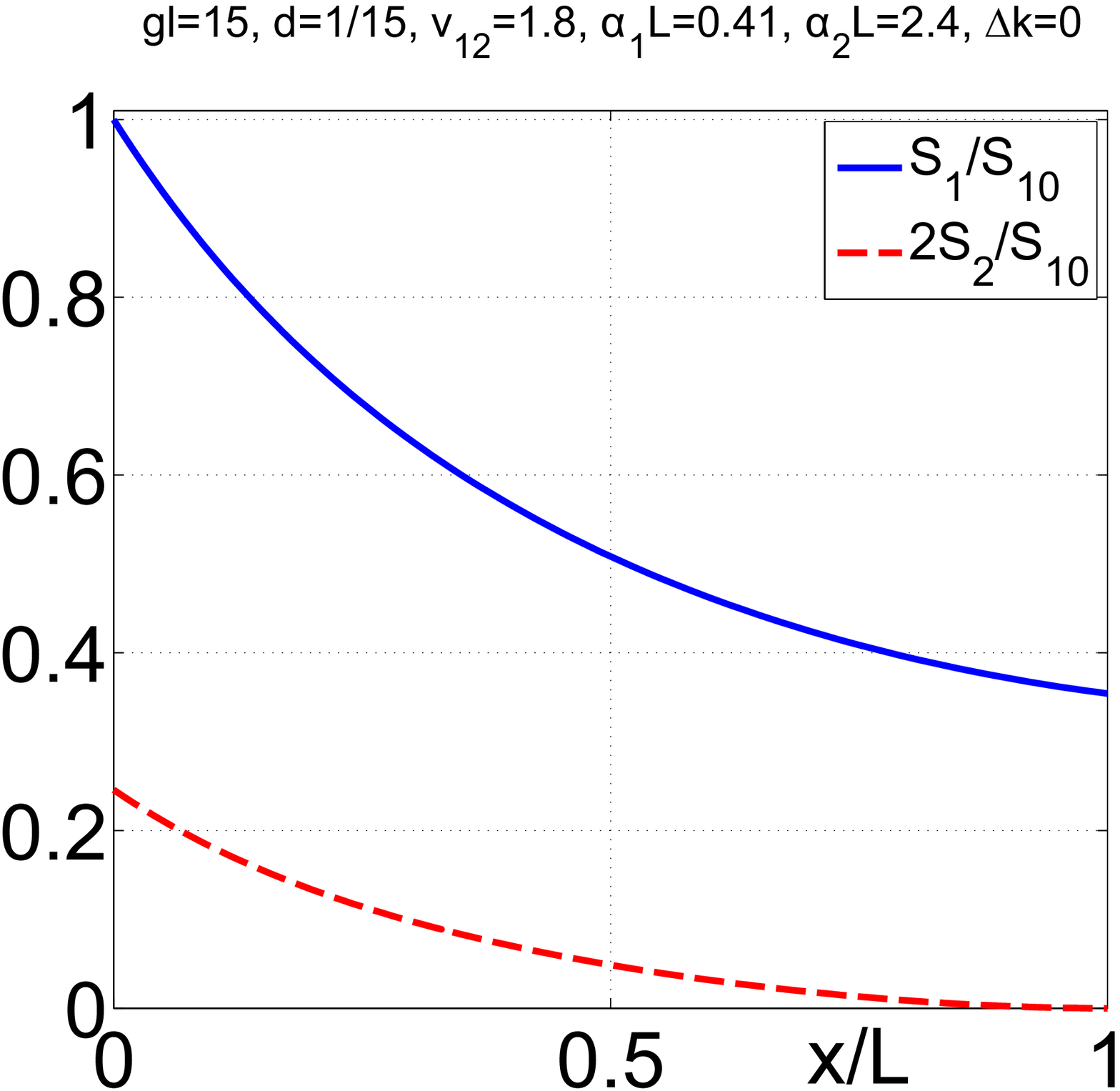}
\includegraphics[width=.49\columnwidth]{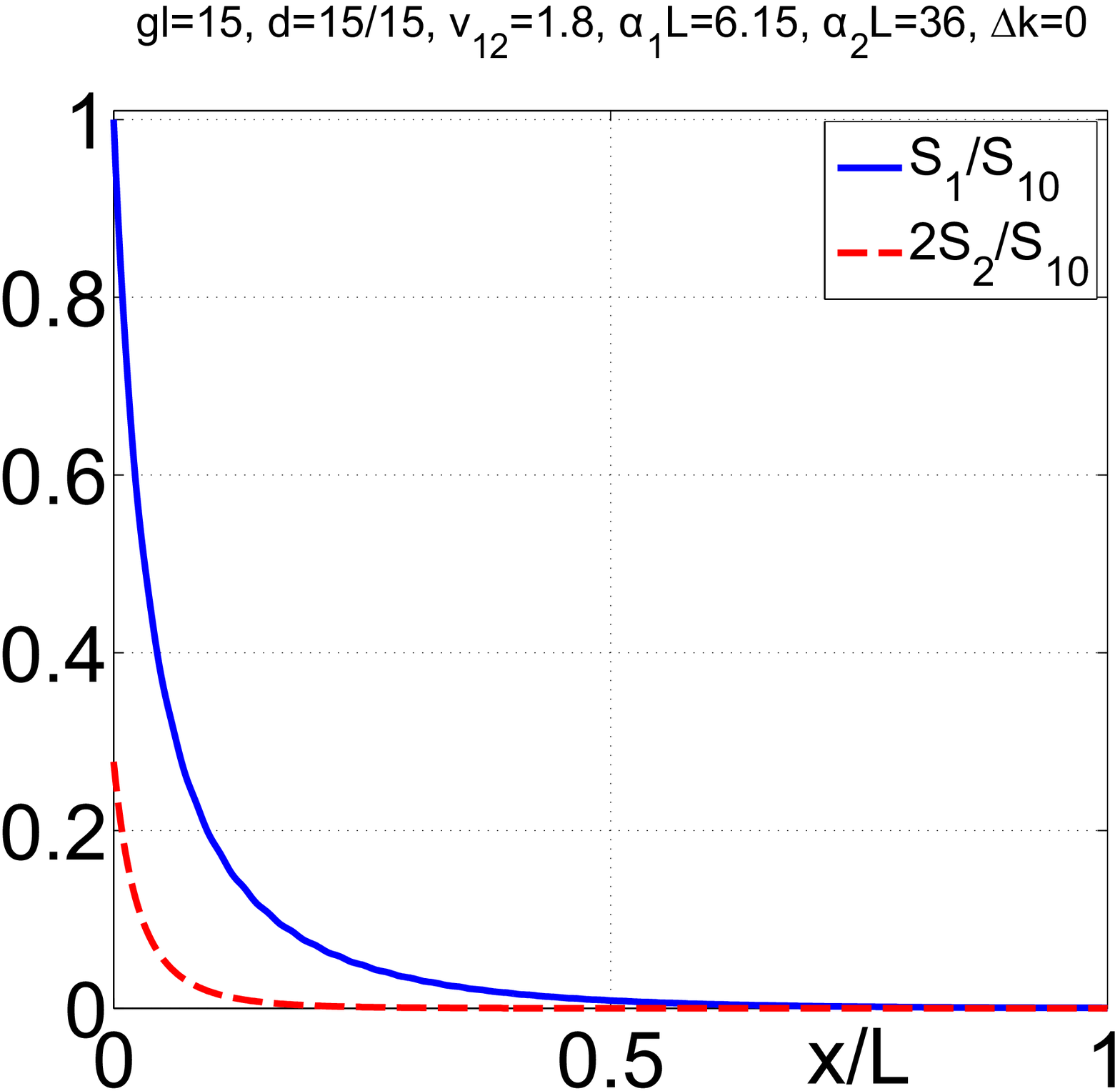} \\
(c)\hspace{0.5\columnwidth} (d)
\caption{BWSHG in a lossy MM:  dependence of energy conversion efficiency on  the metaslab thickness and intensity of the pump pulse. a) and b) $gl=5$, c) and d) $gl=15$. a) and c) $L/l=1/15$, b) and d) $L/l=1$.} \label{nim}
\end{center}
\end{figure}
\subsubsection{Lossy NIM slab}\label{lossy}
Figure~\ref{nim} presents the results of numerical simulations for  energy conversion efficiency at BWSHG with account for the above calculated losses and group velocities pertinent to the given frequencies and the given MM.
\begin{figure}[htbp]
\centering \includegraphics[width=.8\columnwidth]{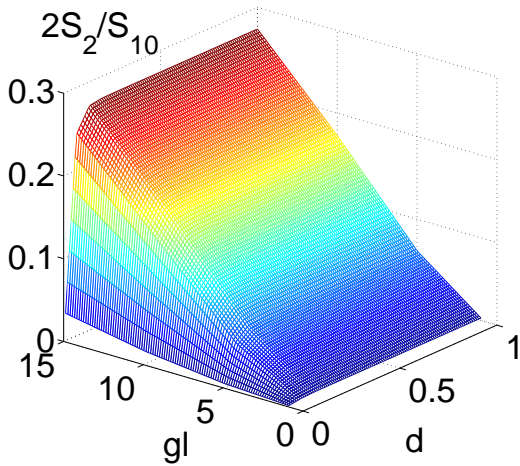}
\caption{BWSHG in a lossy MM: saturation effect in the dependence of  energy conversion efficiency on the metaslab thickness $d$.} \label{3dnim}
\end{figure}

\begin{figure}[htbp]
\centering \includegraphics[width=.49\columnwidth]{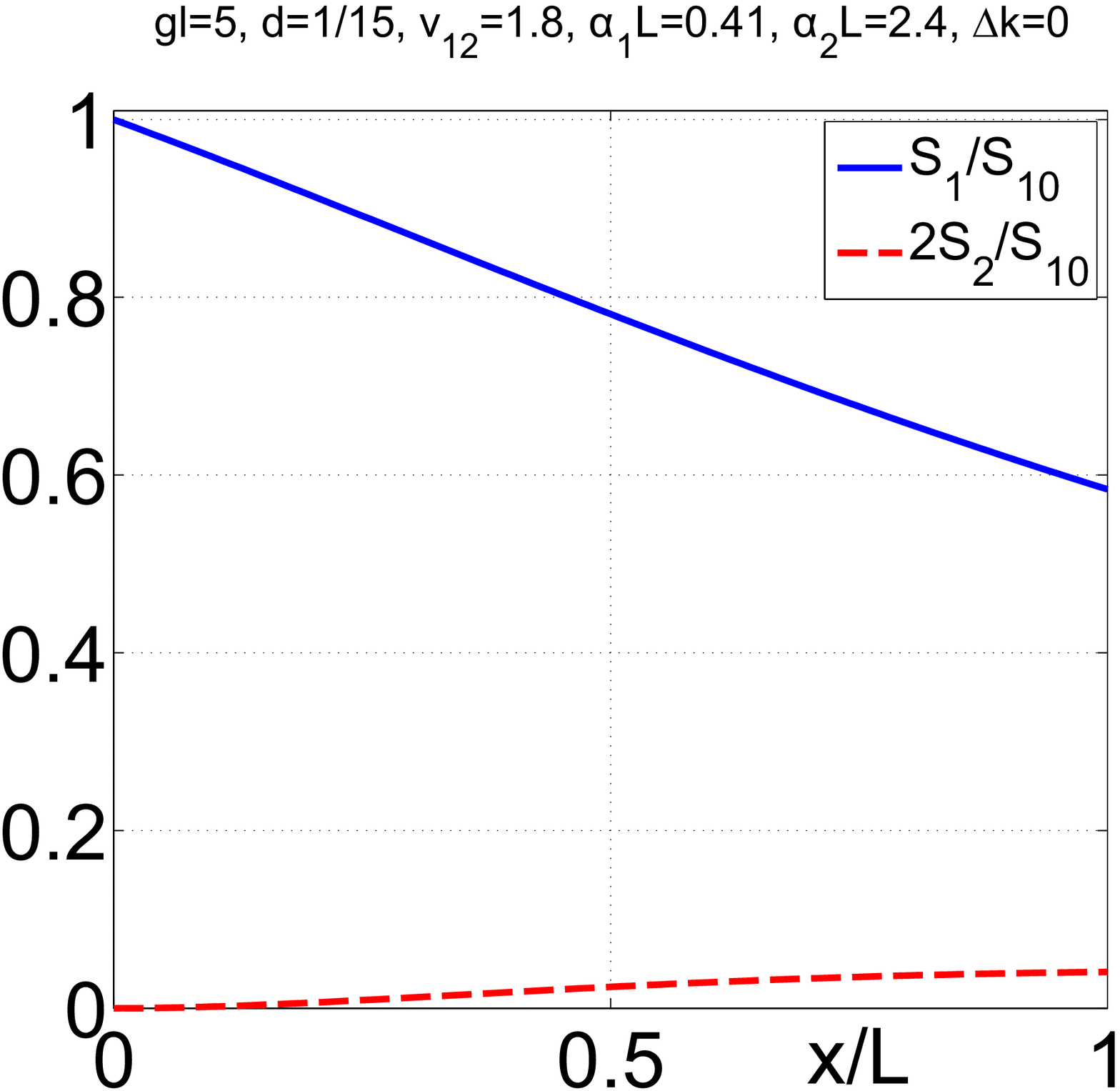}
\includegraphics[width=.49\columnwidth]{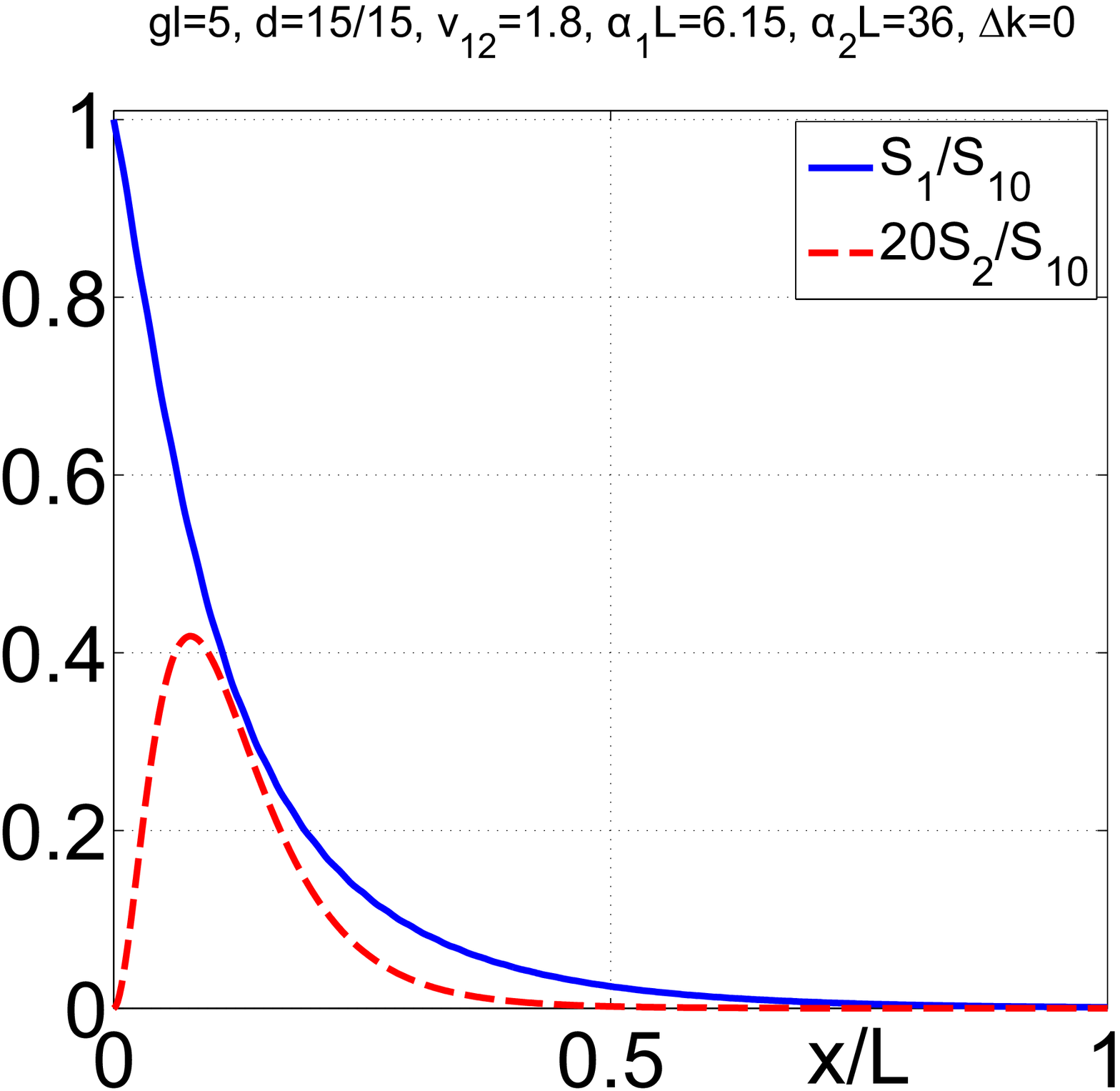}\\
\centering (a)\hspace{0.5\columnwidth}(b)\\
\centering
\includegraphics[width=.49\columnwidth]{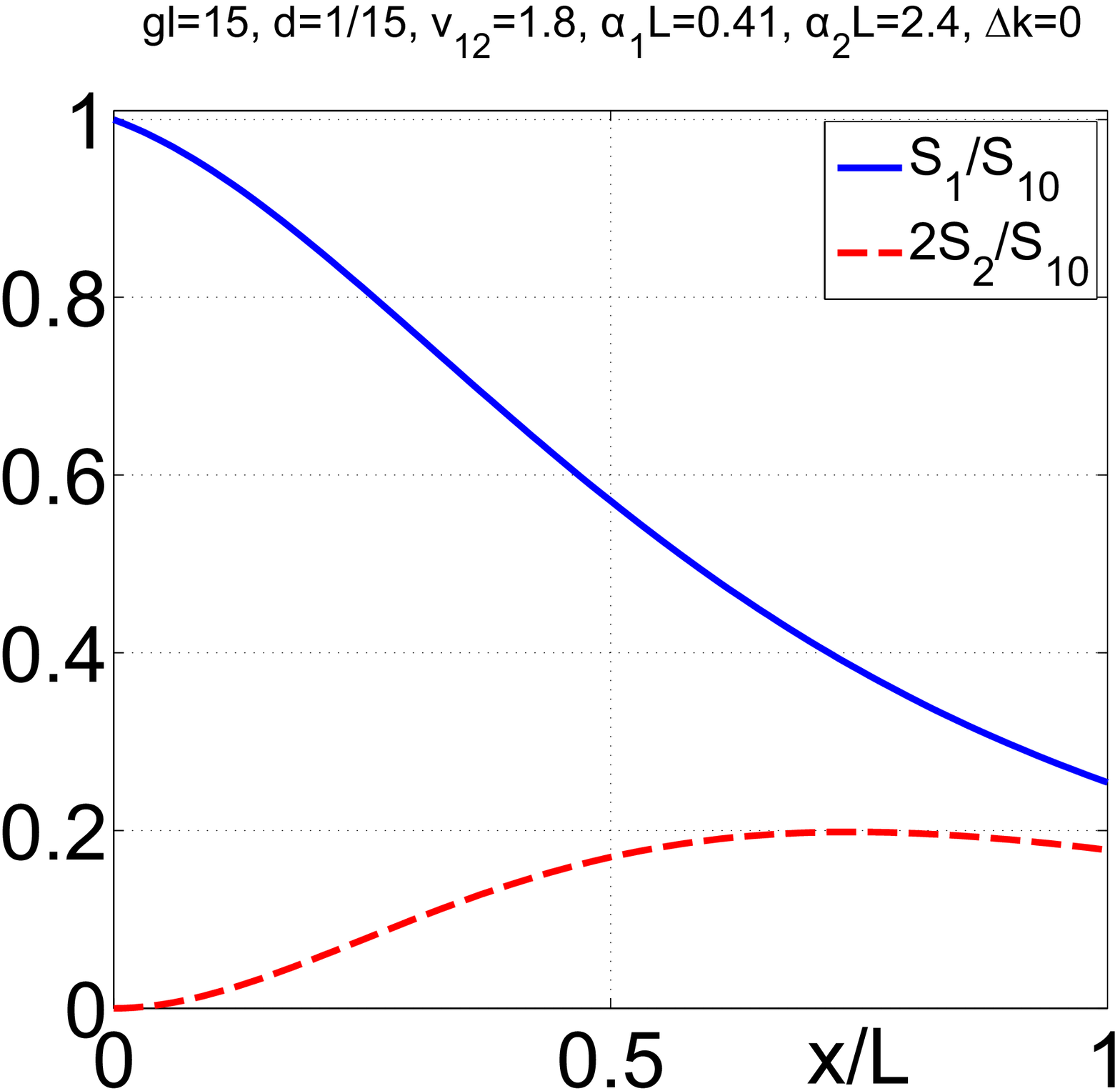}
\includegraphics[width=.49\columnwidth]{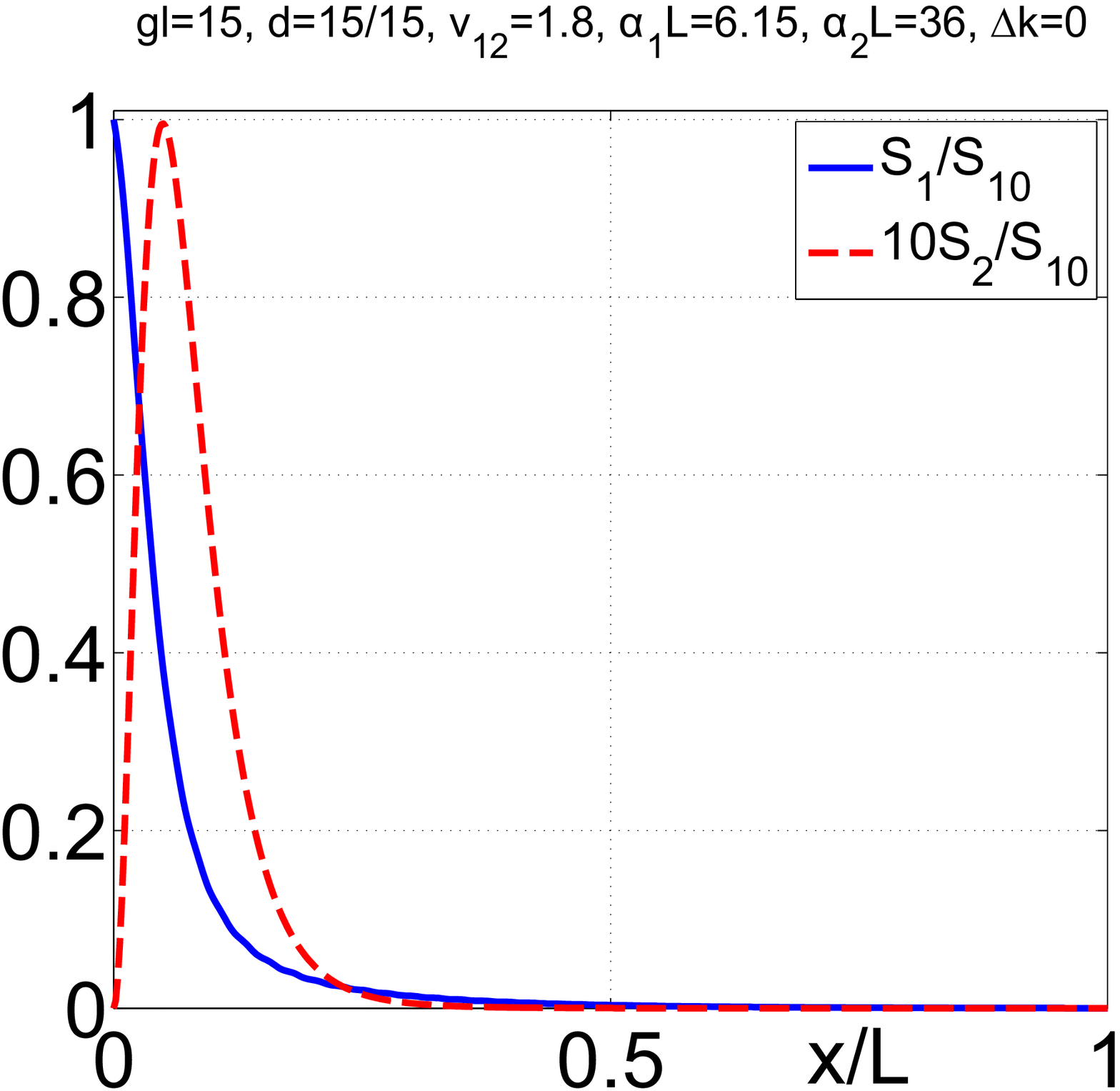}\\
\centering (c)\hspace{0.5\columnwidth} (d)
\caption{Ordinary SHG, all other parameters are the same as in Fig. \ref{nim}.  } \label{pim}
\end{figure}

Figures~\ref{nim} and \ref{3dnim}  show that conversion efficiency grows with increase of  the input pulse amplitude. However, the important \emph{unparalleled} property is that  nonlinear reflectivity rapidly saturates  with increase of the metaslab thickness.   Such unusual behavior is due to backwardness of SH which  propagates against the fundamental beam and is predominantly generated in the area where FH is not yet significantly absorbed. It is seen that quantum nonlinear reflectivity of frequency doubling meta-mirror  may reach the values of about ten percents  at given values of the parameter $gl$. The power reflectivity is two times greater. Such a dependence is in stark contrast with SHG in ordinary materials as seen from comparison with Fig.~\ref{pim}.
It  shows corresponding dependencies in the case of ordinary material with all other parameters the same as in Figs.~\ref{nim}. It is seen that, in the latter case,  SH reaches maximum inside the slab which is due to the interplay of nonlinear conversion and absorption processes. That means the pump strength and the slab thickness must be mutually optimized in this case to maximize the SH output.

\section{Conclusions}
We show the possibility to engineer the metamaterials that satisfy to a set of requirements of paramount  importance for realization of unparallel  nonlinear photonic processes which change photon frequency and propagation direction. The proposed metamaterials enable a set of travelling electromagnetic waves i) which frequencies satisfy to energy conservation low for nonlinear-optical frequency-conversion processes; ii) some of them are extraordinary backward waves with contra-directed energy flux and phase velocity, whereas other(s) are ordinary waves; iii) contra-propagating waves have equal phase velocities, i.e., are phase matched; iv) such properties can be adjusted to different frequencies. Frequency mixing of backward and ordinary waves  possess fundamentally different properties compared to they ordinary counterparts and have important breakthrough applications in photonics. Current mainstream in crafting metamaterials that ensure backward waves relies  on engineering the nanoscopic LC circuits that provide negative magnetic response at optical frequencies. The described in this paper approach is \emph{fundamentally different} and bases on engineering the tailored  coexisting negative and positive dispersion which dictates particular relationship between the frequencies and wavevectors of electromagnetic modes.

Such a general possibility is demonstrated through numerical simulations making use a particular example of the "carbon nanoforest." It is the metamaterial made of carbon nanotubes of particular diameter, height and spacing standing on metallic surface. We show that the  negative and sign-changing dispersions pertinent to such metamaterial can be tailored to support phase matched backward-wave second harmonic generation in the THz and near IR frequency ranges. Losses introduced by the metallic properties of carbon nanotubes and by the particular nano-waveguide modes were investigated and appeared different for the coupled harmonics.  Most practically important, pulsed, regimes of second harmonic generation in such metamaterials was described with the simplified model of  plane travelling waves. A set of coupled partial differential equations  was employed which accounted for dispersion of  group velocities and losses. Since the generated second harmonic  travels in the direction opposite to the fundamental wave, the investigated process  presents a model of the miniature frequency doubling metareflector/metasensor with unparalleled properties.

The described approach can be applied to  engineering the metamaterials that employ tailored positive and negative dispersion to enable extraordinary phase-matched three-wave mixing of contra-propagating waves and their optical parametric amplification which will be described elsewhere.

\section*{Acknowledgements}
This material is based upon work supported in part by the U. S. Army Research Laboratory and the U. S. Army Research Office under grant number W911NF-14-1-0619 and by the Russian Foundation for Basic Research under grant RFBR 15-02-03959A.

\end{document}